\documentclass[twocolumn]{aastex62}
\usepackage{booktabs}

\usepackage{comment}
\usepackage{ulem}
\usepackage{comment}

\received{Jun 28, 2019}
\revised{Dec 20, 2019}
\accepted{Dec 23, 2019, for publication in ApJ}
\shorttitle{Balmer Break Galaxies at $5 \lesssim z \lesssim 8$}
\shortauthors{Mawatari et al.}
\begin{document}

\title{Balmer Break Galaxy Candidates at $z \sim 6$: a Potential View on the Star-Formation Activity at $z \ga 14$}

\correspondingauthor{Ken Mawatari}
\email{mawatari@icrr.u-tokyo.ac.jp}

\author[0000-0003-4985-0201]{Ken Mawatari}
\affiliation{Institute for Cosmic Ray Research, The University of Tokyo, Kashiwa, Chiba 277-8582, Japan}
\affiliation{Department of Environmental Science and Technology, Faculty of Design Technology, Osaka Sangyo University, 3-1-1, Nakagaito, Daito, Osaka, 574-8530, Japan}

\author[0000-0002-7779-8677]{Akio K. Inoue}
\affiliation{Department of Physics, School of Advanced Science and Engineering, Waseda University, 3-4-1, Okubo, Shinjuku, Tokyo 169-8555, Japan}
\affiliation{Waseda Research Institute for Science and Engineering, 3-4-1, Okubo, Shinjuku, Tokyo 169-8555, Japan}
\affiliation{Department of Environmental Science and Technology, Faculty of Design Technology, Osaka Sangyo University, 3-1-1, Nakagaito, Daito, Osaka, 574-8530, Japan}

\author[0000-0002-0898-4038]{Takuya Hashimoto}
\affiliation{Faculty of Science and Engineering, Waseda University, 3-4-1, Okubo, Shinjuku, Tokyo 169-8555, Japan}
\affiliation{Department of Environmental Science and Technology, Faculty of Design Technology, Osaka Sangyo University, 3-1-1, Nakagaito, Daito, Osaka, 574-8530, Japan}
\affiliation{National Astronomical Observatory of Japan, Osawa 2-21-1, Mitaka, Tokyo 181-8588, Japan}

\author[0000-0002-0000-6977]{John Silverman}
\affiliation{Kavli Institute for the Physics and Mathematics of the Universe, The University of Tokyo, Kashiwa, Chiba 277-8583, Japan}

\author[0000-0002-1732-6387]{Masaru Kajisawa}
\affiliation{Research Center for Space and Cosmic Evolution, Ehime University, 2-5 Bunkyo-cho, Matsuyama, Ehime 790-8577, Japan}

\author[0000-0002-7738-5290]{Satoshi Yamanaka}
\affiliation{Waseda Research Institute for Science and Engineering, 3-4-1, Okubo, Shinjuku, Tokyo 169-8555, Japan}
\affiliation{Department of Environmental Science and Technology, Faculty of Design Technology, Osaka Sangyo University, 3-1-1, Nakagaito, Daito, Osaka, 574-8530, Japan}

\author{Toru Yamada}
\affiliation{Institute of Space Astronautical Science, Japan Aerospace Exploration Agency, Sagamihara, Kanagawa 252-5210, Japan}
\affiliation{Astronomical Institute, Tohoku University, Aoba, Aramaki, Aoba-ku, Sendai, Miyagi, 980-8578, Japan}

\author[0000-0002-2951-7519]{Iary Davidzon}
\affiliation{Infrared Processing and Analysis Center, California Institute of Technology, MC 100-22, 770 South Wilson Ave., Pasadena, CA 91125, USA}

\author[0000-0003-3578-6843]{Peter Capak}
\affiliation{Infrared Processing and Analysis Center, California Institute of Technology, MC 100-22, 770 South Wilson Ave., Pasadena, CA 91125, USA}
\affiliation{California Institute of Technology, MC 105-24, 1200 East California Blvd., Pasadena, CA 91125, USA}

\author[0000-0001-7218-7407]{Lihwai Lin}
\affiliation{Institute of Astronomy \& Astrophysics, Academia Sinica, Taipei 106, Taiwan}

\author[0000-0001-5615-4904]{Bau-Ching Hsieh}
\affiliation{Institute of Astronomy \& Astrophysics, Academia Sinica, Taipei 106, Taiwan}

\author{Yoshiaki Taniguchi}
\affiliation{The Open University of Japan, 2-11 Wakaba, Mihama-ku, Chiba 261-8586, Japan}

\author{Masayuki Tanaka}
\affiliation{National Astronomical Observatory of Japan, Osawa 2-21-1, Mitaka, Tokyo 181-8588, Japan}

\author{Yoshiaki Ono}
\affiliation{Institute for Cosmic Ray Research, The University of Tokyo, Kashiwa, Chiba 277-8582, Japan}

\author[0000-0002-6047-430X]{Yuichi Harikane}
\affiliation{Institute for Cosmic Ray Research, The University of Tokyo, Kashiwa, Chiba 277-8582, Japan}
\affiliation{National Astronomical Observatory of Japan, Osawa 2-21-1, Mitaka, Tokyo 181-8588, Japan}

\author[0000-0001-6958-7856]{Yuma Sugahara}
\affiliation{Institute for Cosmic Ray Research, The University of Tokyo, Kashiwa, Chiba 277-8582, Japan}
\affiliation{Department of Physics, Graduate School of Science, The University of Tokyo, 7-3-1 Hongo, Bunkyo, Tokyo, 113-0033, Japan}

\author[0000-0001-7201-5066]{Seiji Fujimoto}
\affiliation{Institute for Cosmic Ray Research, The University of Tokyo, Kashiwa, Chiba 277-8582, Japan}

\author[0000-0002-7402-5441]{Tohru Nagao}
\affiliation{Research Center for Space and Cosmic Evolution, Ehime University, 2-5 Bunkyo-cho, Matsuyama, Ehime 790-8577, Japan}

%\author{}
%\affiliation{}

%% Note that the \and command from previous versions of AASTeX is now
%% depreciated in this version as it is no longer necessary. AASTeX 
%% automatically takes care of all commas and "and"s between authors names.

%% AASTeX 6.2 has the new \collaboration and \nocollaboration commands to
%% provide the collaboration status of a group of authors. These commands 
%% can be used either before or after the list of corresponding authors. The
%% argument for \collaboration is the collaboration identifier. Authors are
%% encouraged to surround collaboration identifiers with ()s. The 
%% \nocollaboration command takes no argument and exists to indicate that
%% the nearby authors are not part of surrounding collaborations.

%% Mark off the abstract in the ``abstract'' environment. 
\begin{abstract}
% Identification

%We report three highest-$z$ passive galaxy candidates.
We search for galaxies with a strong Balmer break (Balmer Break Galaxies; BBGs) at $z \sim 6$ over a 0.41\,deg$^2$ effective area in the COSMOS field. Based on rich imaging data, including data obtained with the Atacama Large Millimeter/submillimeter Array (ALMA), three candidates are identified by their extremely red $K - [3.6]$ colors as well as by non-detection in X-ray, optical, far-infrared (FIR), and radio bands. The non-detection in the deep ALMA observations suggests that they are not dusty galaxies but BBGs at $z \sim 6$, although contamination from Active Galactic Nuclei (AGNs) at $z \sim 0$ cannot be completely ruled out for the moment.
% SED analysis
Our spectral energy distribution (SED) analyses reveal that the BBG candidates at $z \sim 6$ have stellar masses of $\approx 5 \times 10^{10}\,M_{\odot}$ dominated by old stellar populations with ages of $\ga 700$ Myr. 
% SMD 
Assuming that all the three candidates are real BBGs at $z \sim 6$, we estimate the stellar mass density (SMD) to be $2.4^{+2.3}_{-1.3} \times 10^{4}\,M_{\odot}$\,Mpc$^{-3}$. This is consistent with an extrapolation from the lower redshift measurements. 
% SFRD <= should be more emphasized!
The onset of star formation in the three BBG candidates is expected to be several hundred million years before the observed epoch of $z \sim 6$. We estimate the star-formation rate density (SFRD) contributed by progenitors of the BBGs to be 2.4 -- 12 $\times 10^{-5}\,M_{\odot}$\,yr$^{-1}\,$Mpc$^{-3}$ at $z > 14$ (99.7\% confidence range). Our result suggests a smooth evolution of the SFRD beyond $z = 8$. 
\end{abstract}

%% Keywords should appear after the \end{abstract} command. 
%% See the online documentation for the full list of available subject
%% keywords and the rules for their use.

%% From the front matter, we move on to the body of the paper.
%% Sections are demarcated by \section and \subsection, respectively.
%% Observe the use of the LaTeX \label
%% command after the \subsection to give a symbolic KEY to the
%% subsection for cross-referencing in a \ref command.
%% You can use LaTeX's \ref and \label commands to keep track of
%% cross-references to sections, equations, tables, and figures.
%% That way, if you change the order of any elements, LaTeX will
%% automatically renumber them.
%%
%% We recommend that authors also use the natbib \citep
%% and \citet commands to identify citations.  The citations are
%% tied to the reference list via symbolic KEYs. The KEY corresponds
%% to the KEY in the \bibitem in the reference list below. 

\section{Introduction}\label{sec:intro}

Star-formation is the most fundamental process in galaxy formation and evolution. It is important to investigate the cosmic star-formation rate density (SFRD) at a wide redshift range because its evolution can trace cosmic star-formation history (SFH). The SFRD has been measured up to $z = 10$ (e.g., \citealt{Hopkins+06,Oesch+13b,Madau+14,Bouwens+15,Rowan-Robinson+16,Ishigaki+18}). There is a general consensus that the SFRD increases from $z = 0$ to $\sim 2$--$3$ and then monotonically decreases up to $z \sim 8$. The monotonic decrease at $3 < z < 8$ is well expressed by a simple power-law function, $\rho_{\rm SFR} \propto (1 + z)^{\alpha}$, whereas there is a small variation in the slope among different studies \citep{Madau+14,Finkelstein+15a,McLeod+16,Bhatawdekar+19,Oesch+18a}. 

At $z \ga 8$, however, the evolution of the cosmic SFRD is still controversial, which seems to be due to the observational limitations of the current instruments such as the {\it Hubble Space Telescope (HST)}. A smooth evolution from $z \sim 5$ to $\sim 10$ with $-2.6 \leq \alpha \leq -5.8$ is suggested by \citet{Finkelstein+15a,McLeod+16,Bhatawdekar+19} (see also \citealt{Ellis+13} and \citealt{Kikuchihara+19}). In contrast, an accelerated evolution at $z \ga 8$ is suggested by \citet{Bouwens+11a,Oesch+12a,Oesch+14,Oesch+18a}, in which the power-law slope dramatically changes from $\alpha \approx -4$ at $z < 8$ to $\alpha = -10.9$ at $8< z \lesssim 10$. \citet{Oesch+18a} claim that the rapid decline of SFRD beyond $z \approx 8$ is naturally explained by the number density evolution of dark matter halos. This explanation is also supported by \citet{Harikane+18a} who reproduced the rapid SFRD decrease assuming no redshift dependence on a tight relation among the halo mass, SFR, and dark matter accretion rate. To obtain a definitive conclusion, significant improvements on the SFRD measurements at $z>8$ are required.

If galaxies experience passive evolution with no or little star-formation activity for more than several hundreds of million years, their spectra are dominated by A-type or cooler stars with a Balmer/4000\,\AA\ break (e.g., \citealt{Leitherer+99,Wiklind+08}). Studying passive or Balmer Break Galaxies (BBGs; \citealt{Wiklind+08}) at high redshift can potentially help explore a redshift frontier of cosmic SFH because such galaxies should have undergone intense star-formation a long time before they are observed. For example, a spectroscopically confirmed galaxy at $z = 9.1$ has a strong Balmer break \citep{Hashimoto+18a}, whereas this galaxy also shows current star-formation and cannot be regarded as a pure passive galaxy. \citet{Hashimoto+18a} analyzed the spectral energy distribution (SED), concluding that the galaxy started star-formation at redshift as high as $z\sim15$. Because the $z = 9.1$ galaxy was first selected by the standard Lyman break technique and its Balmer break was found serendipitously, we cannot obtain any statistical quantities for such galaxies showing the Balmer break. A systematic survey of BBGs at high-$z$, namely $z > 5$, is significantly interesting to investigate the cosmic star formation at $z > 10$.

Passive galaxies have been well investigated at $z < 3$, which reveal that the number density of the passive galaxies decreases with increasing redshift \citep{Kajisawa+11b,Dominguez+11,Muzzin+13c,Davidzon+17}. At $z > 3$, the Balmer break is shifted to $\lambda > 1.5\,\mu$m, making detection of passive galaxies challenging. The redshift record of spectroscopically confirmed passive galaxies reaches as large as $z \approx 4$ \citep{Glazebrook+17,Tanaka+19}. Even at larger redshifts of $4 \lesssim z \lesssim 6$, the number of photometric BBG candidates increases because of the extremely deep and wide near-infrared (NIR) survey data \citep{Rodighiero+07,Wiklind+08,Mancini+09,JSHuang+11,Caputi+12,Nayyeri+14,Santini+19,Merlin+19}. \citet{Mawatari+16b} have proposed a color selection scheme to isolate BBGs at $5 \lesssim z \lesssim 8$ and identified three candidates in the $Spitzer$ Extended Deep Survey (SEDS; \citealt{Ashby+13}) UDS field. 

In most previous studies on photometric identification of high-$z$ BBGs, there still remains possible contamination from dusty galaxies with a similar red rest-frame optical color to the BBGs (e.g., \citealt{Brammer+09}). This is due to the lack of a sufficiently deep constraint on dust emission in the mid-infrared (MIR) to far-infrared (FIR) regions. The passive galaxy at $z \sim 6.5$ reported in the pioneering work by \citet{Mobasher+05} was later identified as a dusty contaminant at $z < 3$ detected with $Spitzer$/MIPS \citep{Dunlop+07}. One of the three BBG candidates reported by \citet{Mawatari+16b} (the object ID: SEDS\_UDS\_BBG-34) was found to be a low-$z$ dusty galaxy because of detections in the new FIR imaging data from the SCUBA-2 Cosmology Legacy Survey (S2CLS; \citealt{Geach+17}) and the Atacama Large Millimeter/submillimeter Array (ALMA) SCUBA-2 UDS survey (AS2UDS; \citealt{Stach+19}). ALMA may be the only instrument that can offer critical data to resolve the degeneracy between passive and dusty red galaxies because of its unprecedented sensitivity and spatial resolution \citep{Schreiber+18,Santini+19}. 

In this work, we apply a color selection of BBGs to the deep and wide-area imaging data available in an effective area of $0.41$\,deg$^2$ in the COSMOS field. We further conduct ALMA observations to remove contamination from dusty galaxies. This study is structured as follows. The imaging data used in this work are summarized in Section~2. Spectral templates of galaxies and AGNs to tune the color selection criteria and to analyze the SEDs are described in Section~3. In Sections~4 and 5, we describe the selection of $5 \lesssim z \lesssim 8$ BBG candidates and follow-up observations with ALMA. In Sections~6 and 7, we discuss the sample significance through SED analyses and implications on their progenitors' SFRD. We use the AB magnitude system \citep{OkeGunn83} and adopt a cosmology with $H_{0}=70.4$ km s$^{-1}$ Mpc$^{-1}$, $\Omega_{M}=0.272$, and $\Omega_{\Lambda}=0.728$ \citep{Komatsu+11}.

\section{Multi-band imaging data}\label{sec:data}

\startlongtable
\begin{deluxetable*}{lccccc}
\tablecaption{COSMOS dataset \label{tb:data_sum}}
\tablehead{
\colhead{Instrument} & \colhead{Filter} & \colhead{FWHM\tablenotemark{a}} & \colhead{Limiting flux\tablenotemark{b}} & \colhead{Survey} & \colhead{Reference\tablenotemark{c}}\\
\colhead{} & \colhead{} & \colhead{(arcsec)} & \colhead{} & \colhead{} & \colhead{}
}
\colnumbers
\startdata
$HST$/ACS & $F814W$ & 0.10 & 27.5(26.6\tablenotemark{d})\,mag & $HST$-COSMOS & (1)(2) \\
%Subaru/HSC & $g$ & 0.75 & 26.9\,mag & HSC-SSP/UD S16A & (3) \\
Subaru/HSC & $g$ & 0.79 & 27.3\,mag & HSC-SSP/UD S18A & (3) \\
%Subaru/HSC & $r$ & 0.54 & 27.0\,mag & HSC-SSP/UD S16A & (3) \\
Subaru/HSC & $r$ & 0.66 & 27.3\,mag & HSC-SSP/UD S18A & (3) \\
%Subaru/HSC & $i$ & 0.61 & 26.5\,mag & HSC-SSP/UD S16A & (3) \\
Subaru/HSC & $i$ & 0.64 & 27.1\,mag & HSC-SSP/UD S18A & (3) \\
%Subaru/HSC & $z$ & 0.56 & 26.2\,mag & HSC-SSP/UD S16A & (3) \\
Subaru/HSC & $z$ & 0.58 & 26.8\,mag & HSC-SSP/UD S18A & (3) \\
%Subaru/HSC & $y$ & 0.70 & 25.2\,mag & HSC-SSP/UD S16A & (3) \\
Subaru/HSC & $y$ & 0.70 & 25.8\,mag & HSC-SSP/UD S18A & (3) \\
%Subaru/HSC & $NB816$ &  & ($<$26.5) & HSC-SSP & proposal \\
%Subaru/HSC & $NB921$ &  & ($<$26.2) & HSC-SSP & proposal \\
%Subaru/HSC & $NB101$ &  & ($<$24.8) & HSC-SSP & proposal \\
VISTA/VIRCAM & $Y$ & 0.8 & 25.8\,mag & UltraVISTA/UD DR3 & (4) \\
VISTA/VIRCAM & $J$ & 0.77 & 25.7\,mag & UltraVISTA/UD DR3 & (4) \\
VISTA/VIRCAM & $H$ & 0.75 & 25.5\,mag & UltraVISTA/UD DR3 & (4) \\
VISTA/VIRCAM & $K_s$ & 0.75 & 25.2\,mag & UltraVISTA/UD DR3 & (4) \\
%Mayall/NEWFIRM
$Spitzer$/IRAC & 3.6$\mu$m & 1.7 & 23.9\,mag & SPLASH & (5)(6)\\
$Spitzer$/IRAC & 4.5$\mu$m & 1.6 & 24.0\,mag & SPLASH & (5)(6)\\
$Spitzer$/IRAC & 5.8$\mu$m & 1.8 & 20.8\,mag  & S-COSMOS & (7)\\
$Spitzer$/IRAC & 8.0$\mu$m & 2.1 & 20.7\,mag  & S-COSMOS & (7)\\
$Spitzer$/MIPS & 24$\mu$m & 5.9 & 19.0\,mag & S-COSMOS & (8)\\
$Spitzer$/MIPS & 70$\mu$m & 18.6 & 14.1\,mag & S-COSMOS & (9)\\
$Herschel$/PACS & 100$\mu$m & 7.2 & 14.2\,mag & PEP DR1 & (10)\\
$Herschel$/PACS & 160$\mu$m & 12.0 & 13.4\,mag & PEP DR1 & (10)\\
$Herschel$/SPIRE & 250$\mu$m & 18.15 & 12.9\,mag & HerMES DR4 & (11)(12) \\
$Herschel$/SPIRE & 350$\mu$m & 25.15 & 12.6\,mag & HerMES DR4 & (11)(12) \\
$Herschel$/SPIRE & 500$\mu$m & 36.3 & 12.7\,mag & HerMES DR4 & (11)(12) \\
JCMT/SCUBA-2 & 850$\mu$m & 8.0 & 14.1\,mag & S2CLS & (13) \\
VLA & 1.4\,GHz & $1.5 \times 1.4$ & 75\,$\mu$Jy & VLA-COSMOS/Large & (14) \\
VLA & 3\,GHz & $0.75$ & 11.5\,$\mu$Jy & VLA-COSMOS/Large & (15) \\
$XMM-Newton$ & 0.5-2\,keV & -- & $1.0 \times 10^{-15}$\,erg\,cm$^{-2}$\,s$^{-1}$ & XMM-COSMOS & (16) \\
$XMM-Newton$ & 2-10\,keV & -- & $5.6 \times 10^{-15}$\,erg\,cm$^{-2}$\,s$^{-1}$ & XMM-COSMOS & (16) \\
$XMM-Newton$ & 5-10\,keV & -- & $1.1 \times 10^{-14}$\,erg\,cm$^{-2}$\,s$^{-1}$ & XMM-COSMOS & (16) \\
$Chandra$ & 0.5-2\,keV & -- & $4.9 \times 10^{-16}$\,erg\,cm$^{-2}$\,s$^{-1}$ & Chandra-COSMOS Legacy & (17) \\
$Chandra$ & 2-10\,keV & -- & $3.1 \times 10^{-15}$\,erg\,cm$^{-2}$\,s$^{-1}$ & Chandra-COSMOS Legacy & (17) \\
$Chandra$ & 0.5-10\,keV & -- & $1.9 \times 10^{-15}$\,erg\,cm$^{-2}$\,s$^{-1}$ & Chandra-COSMOS Legacy & (17) \\
\enddata
\tablenotetext{a}{FWHMs are estimated by stacking 100 -- 200 bright stars in the survey area for $HST$, Subaru/HSC, VISTA/VIRCAM, and $Spitzer$/IRAC images. FWHMs for $Spitzer$/MIPS and $Herschel$ are quoted from reference literature \citep{Lutz+11,Andrews+17}. The instrumental beam size is shown for the JCMT/SCUBA-2 image.}
\tablenotetext{b}{$5\sigma$ limiting magnitudes are measured adopting 2 $\times$ PSF (FWHM) diameter apertures and an aperture correction for $HST$, Subaru/HSC, VISTA/VIRCAM, and $Spitzer$/IRAC images. For $Spitzer$/MIPS, $Herschel$/PACS, JCMT/SCUBA-2, and VLA images, $5 \sigma$ limiting magnitudes or flux densities from the references are shown. For the $Herschel$/SPIRE images that severely suffer from source confusion, the confusion limits estimated in \citet{Oliver+12} are listed. For $XMM-Newton$ and $Chandra$ data, limiting fluxes achieved over $50$\,\% of the survey area are adopted from the references.}
\tablenotetext{c}{(1)~\citet{Koekemoer+07}, (2)~\citet{Scoville+07}, (3)~\citet{Aihara+19}, (4)~\citet{McCracken+12}, (5)~P.I.: P.~Capak, (6)~\citet{Laigle+16}, (7)~\citet{Sanders+07},  (8)~\citet{LeFloch+09}, (9)~\citet{Frayer+09}, (10)~\citet{Lutz+11}, (11)~\citet{Oliver+12},  (12)~\citet{Andrews+17}, (13)~\citet{Geach+17}, (14)~\citet{Schinnerer+07}, (15)~\citet{Smolcic+17}, (16)~\citet{Cappelluti+09}, (17)~\citet{Civano+16}}
\tablenotetext{d}{A $5\sigma$ limiting magnitude estimated with 0.6\,arcsec diameter apertures which are actually used in this work.}
%\tablecomments{Note that }
\end{deluxetable*}

\begin{figure}[]
\begin{center}
\includegraphics[width=1.0\linewidth, angle=0]{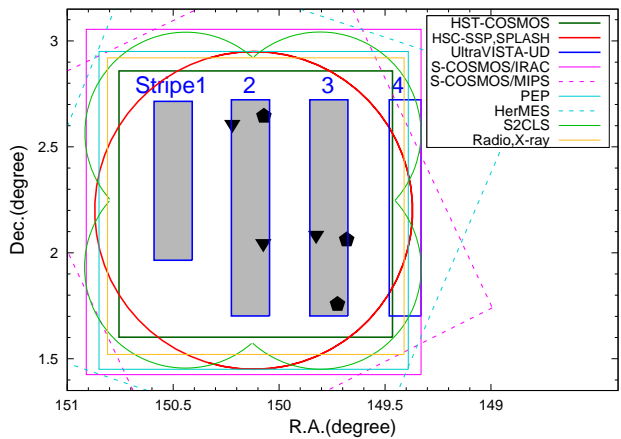}
\end{center}
\caption{Sky coverage of the survey data used in this work: $HST$-COSMOS (dark green box), HSC-SSP UD and SPLASH (red circle), UltraVISTA UD (blue boxes), S-COSMOS (magenta solid box for the IRAC $5.8$ and $8.0\,\mu$m, and magenta dashed box for the MIPS $24$ and $70\,\mu$m), PEP (cyan solid box), HerMES (cyan dashed box), S2CLS (light green curve), VLA-COSMOS, $XMM$-COSMOS, and $Chandra$-COSMOS Legacy (yellow box). The blue colored numbers are assigned for the four UltraVISTA UD stripes. The area where we searched for BBGs at $5 \lesssim z \lesssim 8$ is emphasized by the gray color. The observed BBG candidates with and without ALMA detection (see \S\ref{sec:ALMA}) are shown by filled triangles and pentagons, respectively. \label{fig:coverage}}
\end{figure}

We gathered multi-band photometric data available in the COSMOS field. In Table~\ref{tb:data_sum}, we present the instruments, filters, full-width-at-half-maximum (FWHM) of the point spread function (PSF), limiting flux, survey name, and references. Figure~\ref{fig:coverage} shows the spatial coverage of each data. In the following section, we explain each data set. 

We use deep $Spitzer$'s Infrared Array Camera (IRAC: \citealt{Fazio+04}) images at a wavelength of $3.6\,\mu$m and $4.5\,\mu$m from the $Spitzer$ Large Area survey with Hyper-Suprime-Cam (SPLASH; P. I. is P. Capak; \citealt{Laigle+16}). Ground-based $Y$-, $J$-, $H$-, and $K$-band images were also available from the UltraVISTA survey \citep{McCracken+12}. We used only the deepest data in the four Ultra-Deep (UD) stripes that are included in the data release 3 (DR3\footnote{http://ultravista.org/release3/}). We call these stripes UVISTA UD Stripe1, 2, 3, and 4 (Figure~\ref{fig:coverage}). These stripes cover $\sim 0.66$\,deg$^2$ in the SPLASH field. 

There are {\it HST} Advanced Camera for Surveys (ACS) $F814W$ imaging data taken in the original COSMOS $HST$ Treasury project (hereafter, $HST$-COSMOS; \citealt{Scoville+07,Koekemoer+07}). The $F814W$-band image and catalog were downloaded from the COSMOS website\footnote{http://cosmos.astro.caltech.edu/page/hst}. The $HST$-COSMOS data cover the UVISTA UD stripes, except for the westernmost Stripe4. Other optical imaging data at $g$-, $r$-, $i$-, $z$-, and $y$-bands are available from the Subaru strategic program using Hyper Suprime-Cam (HSC-SSP; \citealt{Aihara+18a,Miyazaki+18,Komiyama+18,Kawanomoto+18,Furusawa+18}). We used the HSC-SSP public data (PDR2 or S18A; \citealt{Aihara+19}) in their deepest UltraDeep (UD) layer that covers the three UVISTA UD stripes 1 to 3. 

At MIR wavelength range, IRAC/5.8-, 8.0-, MIPS/24-, and 70\,$\mu$m-bands imaging data are available from the $Spitzer$ COSMOS survey (S-COSMOS; \citealt{Sanders+07,LeFloch+09,Frayer+09}). These $Spitzer$ MIR data\footnote{http://cosmos.astro.caltech.edu/page/spitzer} cover the all four UVISTA UD stripes, while the depth in the westernmost UVISTA UD Stripe4 is shallow. 

There are two major FIR surveys conducted by ESA's $Herschel$ space observatory \citep{Pilbratt+10}. One is the Photodetector Array Camera and Spectrometer (PACS; \citealt{Poglitsch+10}) Evolutionary Probe (PEP; \citealt{Lutz+11}) survey. We used the $100$ and $160\,\mu$m-band images and catalogs from the PEP first data release (DR1\footnote{http://www.mpe.mpg.de/ir/Research/PEP/DR1}). These images cover the four UVISTA UD stripes, except for a part of the westernmost Stripe4. Another survey is the $Herschel$ Multi-Tiered Extragalactic Survey (HerMES; \citealt{Oliver+12}), which uses the Spectral and Photometric Imaging Receiver (SPIRE; \citealt{Griffin+10}) at wavelengths of $250$, $350$, and $500\,\mu$m. We obtained the SPIRE images and catalogs in the COSMOS field from the fourth HerMES data release (DR4), through the $Herschel$ Database in Marseille (HeDaM\footnote{http://hedam.lam.fr/HerMES/index/dr4}) operated by CeSAM and hosted by the Laboratoire d'Astrophysique de Marseille. The Submillimeter Common-User Bolometer Array-2 (SCUBA-2; \citealt{Holland+13}) equipped with the James Clerk Maxwell Telescope (JCMT) provides similarly deep FIR images to the {\it Herschel} images but better spatial resolution. We used the SCUBA-2 $850\,\mu$m map of the COSMOS field taken as a part of the SCUBA-2 Cosmology Legacy Survey (S2CLS\footnote{http://dx.doi.org/10.5281/zenodo.57792}; \citealt{Geach+17}). This S2CLS $850\,\mu$m map covers the four UVISTA UD stripes, except for a part of the westernmost Stripe4. We note that the S2CLS $850\,\mu$m map is inhomogeneous and the easternmost UVISTA UD Stripe1 falls into a shallow area. Fortunately, this does not affect this work because no BBG candidate is found there. 

There are rich radio and X-ray data available in the COSMOS field. Homogeneously deep $1.4\,$GHz and $3\,$GHz maps were provided by the Karl G. Jansky Very Large Array (VLA) COSMOS Large project \citep{Schinnerer+07,Smolcic+17}. We used the catalogs\footnote{http://cosmos.astro.caltech.edu/page/radio} of $\sim 2,400$ and $\sim 11,000$ sources with a signal-to-noise ratio larger than five ($S/N$ $> 5$) in the $1.4\,$GHz and $3\,$GHz maps, respectively. In X-ray, both $XMM-Newton$ and $Chandra$ satellites observed the $\sim 2$\,deg$^2$ area in the $XMM$-COSMOS survey \citep{Cappelluti+09} and $Chandra$-COSMOS Legacy survey \citep{Civano+16}. We used the $XMM$-COSMOS catalog of $\sim 1,800$ sources and the $Chandra$-COSMOS Legacy catalog of $\sim 4,000$ sources from the COSMOS website\footnote{http://cosmos.astro.caltech.edu/page/xray}. 

In summary, we analyzed the UVISTA UD Stripe1, 2, and 3, where the above multi-wavelength imaging data were homogeneously deep.

\section{Template colors}

Because the Balmer break at $z > 5$ is redshifted to an observed wavelength longer than the $K$-band, the main color selection criterion is red $K-$[3.6] to capture the break. In addition, we set a secondary color selection criterion of [3.6]$-$[4.5] to reject the dusty galaxies showing similar red $K - [3.6]$ colors. To make a suitable set of the color selection criteria for $z\sim6$ BBGs, \citet{Mawatari+16b} investigated the galaxy colors on the $K-$[3.6] versus [3.6]$-$[4.5] two color diagram with SED model templates. In this section, we present an updated set of the color selection criteria for $z\sim6$ BBGs based on an expanded analysis of the model template colors.

\subsection{Galaxy Model: Star + Nebular + Dust}\label{sec:modelgal}

Our galaxy SED models consist of three components: the stellar continuum models \citep{BruzualCharlot03}, the nebular emission models \citep{Inoue+11b}, and the empirical dust emission templates \citep{Rieke+09}. We call them ``Star+Nebular+Dust''.
These models were also used in our previous studies \citep{Hashimoto+18a,Hashimoto+19a,Tamura+19}. 
For the stellar continuum models, we assume a Chabrier initial mass function (IMF) \citep{Chabrier03} with lower and upper mass cutoffs of 0.1 and 100 $M_{\odot}$, respectively. The SFH is assumed to be a constant SFR or is assumed to be exponentially declining/rising with various e-folding timescales. The parameter ranges for age ($T_{\rm age}$), metallicity ($Z$), and e-folding timescale ($\tau_{\rm SFH}$) are as follows: $1$\,Myr $< T_{\rm age} <$ the cosmic age at a given redshift, $0.0001 < Z < 0.02$, $0.01$\,Gyr $\leq \tau_{\rm SFH} \leq 10$\,Gyr. 
%The number of the Star+Nebular+Dust templates is 1,575 before applying dust attenuation and redshift. 

The nebular continuum and emission line fluxes are calculated from the ionizing photon production rate and metallicity of the stellar components \citep{BruzualCharlot03} in the same manner as \citet{Inoue+11b}. The escape fraction of ionizing photons is assumed to be zero. Fluxes at rest-frame wavelengths equal to or shorter than Ly$\alpha$ are attenuated by the intergalactic medium (IGM), following the analytic model of \citet{Inoue+14}. For dust attenuation, $A_V$, we apply the same amount to both nebular and stellar continua, assuming the Calzetti law \citep{Calzetti+00}. We also put the maximum $A_V$ as a function of SFR, $A_V < {\rm max}(4 \times {\rm SFR}^{0.3}, 3.5)$, to avoid the templates of extremely dusty and passive galaxies never observed so far (see discussion in Appendix~\ref{sec:Av2SFR}). 

The energy attenuated by dust is re-radiated in the infrared wavelengths ($5\,\mu$m $\lesssim \lambda \lesssim$ $1000\,\mu$m). The dust emission is described by empirical templates of nearby infrared-bright galaxies \citep{Rieke+09} as a function of the total infrared luminosity ($L_{\rm IR}$). We selected the template with $L_{\rm IR}$ equal to the luminosity attenuated by dust.

\subsection{AGN templates}\label{sec:tempAGN}

Our AGN template set consists of nine empirical and 24,000 theoretical spectra. The empirical templates are taken from the SWIRE template library \citep{Polletta+07}: three type-1 AGNs (``QSO1'', ``TQSO1'', and ``BQSO1''), four type-2 AGNs (``Sey2'', ``Sey1.8'', ``QSO2'', and ``Torus''), and two starburst galaxies with AGNs (``Mrk231'' and ``I19254''). Because all of them are constructed based on various types of observed AGNs, their spectra include the host galaxy contribution. 

The theoretical AGN templates were constructed by \citet{Fritz+06} and \citet{Feltre+12} based on comprehensive modeling of a dusty torus around a black hole (BH). Their model (hereafter, ``TORUS'') has realistic torus geometry parameterized by an outer-to-inner radial ratio, an opening angle, gas density profile, optical depth at equatorial plane, and a viewing angle. Following the unified AGN picture \citep{Antonucci+93,UrryPadovani95}, we consider type-1 and type-2 AGNs depending on the viewing angle. \citet{Fritz+06} mentioned that not only the torus emission but also contribution from the host-galaxy are needed to reproduce the actual observed SEDs of AGNs, except for a few cases of type-1 AGNs (see also discussion in \S\ref{sec:disc_z0AGN}).

\subsection{Expected colors for $z \sim 6$ BBGs}\label{sec:tempC}

\begin{figure*}[]
\begin{center}
\includegraphics[width=1.0\linewidth, angle=0]{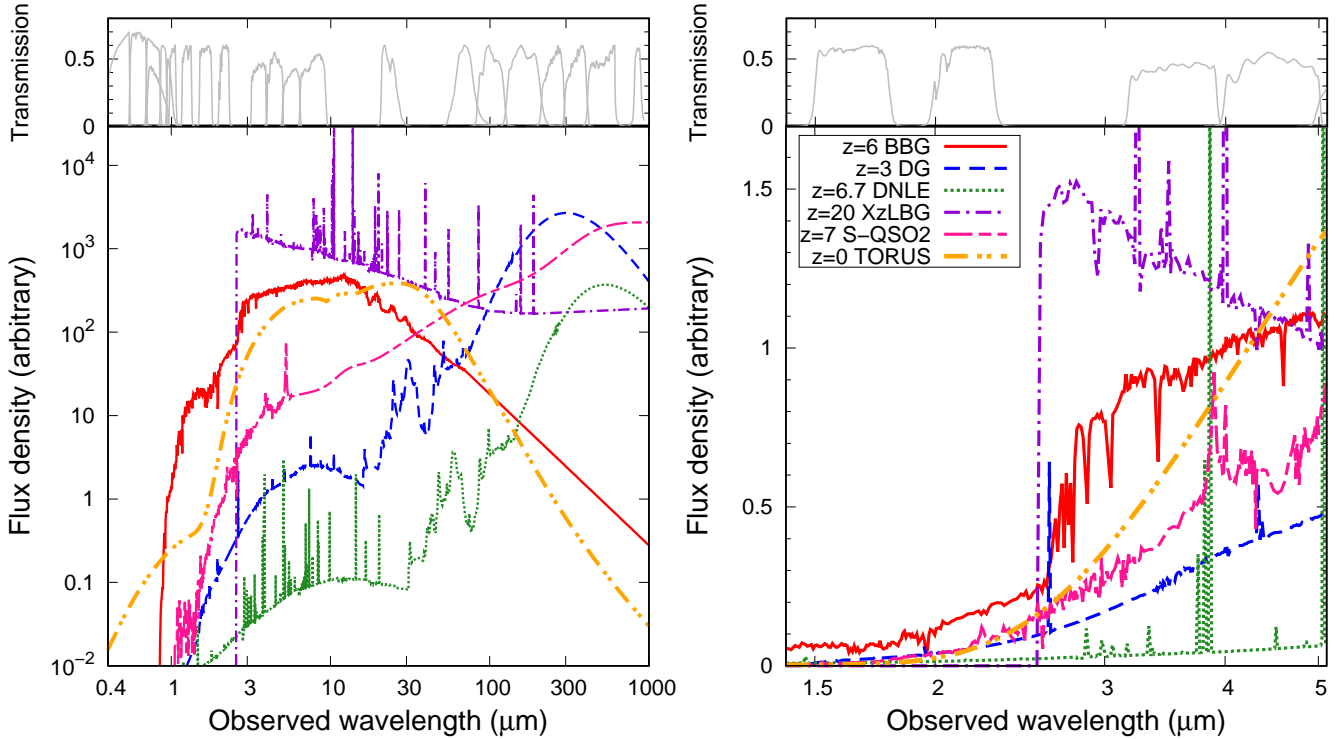}
\end{center}
\caption{Example spectra of the galaxy and AGN templates with extremely red $K - [3.6]$ but flat $[3.6] - [4.5]$ colors. The left and right panels show the same spectra, whereas in the right panel, the flux is in the linear scale and the wavelength range is limited to the NIR regime. Four types of galaxies come from the Star+Nebular+Dust model library: a passive galaxy at $z = 6$ with $A_V = 0$ (BBG), a dusty galaxy at $z = 3$ with $A_V =4$ (DG), an extremely young and dusty galaxy with strong nebular emission lines at $z = 6.7$ with $A_V =2$ (DNLE), and an LBG at $z = 20$ with $A_V = 0$ (XzLBG). Two AGN spectra come from the empirical and theoretical AGN template library: a type-2 QSO template of the SWIRE template library at $z = 7$ (S-QSO2) and a heavily obscured dust torus model at $z = 0$ (TORUS). We arbitrarily scaled the individual spectra for display purposes. The filter response curves used in this work are shown in the top panels. \label{fig:modelspec}}
\end{figure*}

\begin{figure*}[]
\begin{center}
\includegraphics[width=1.0\linewidth, angle=0]{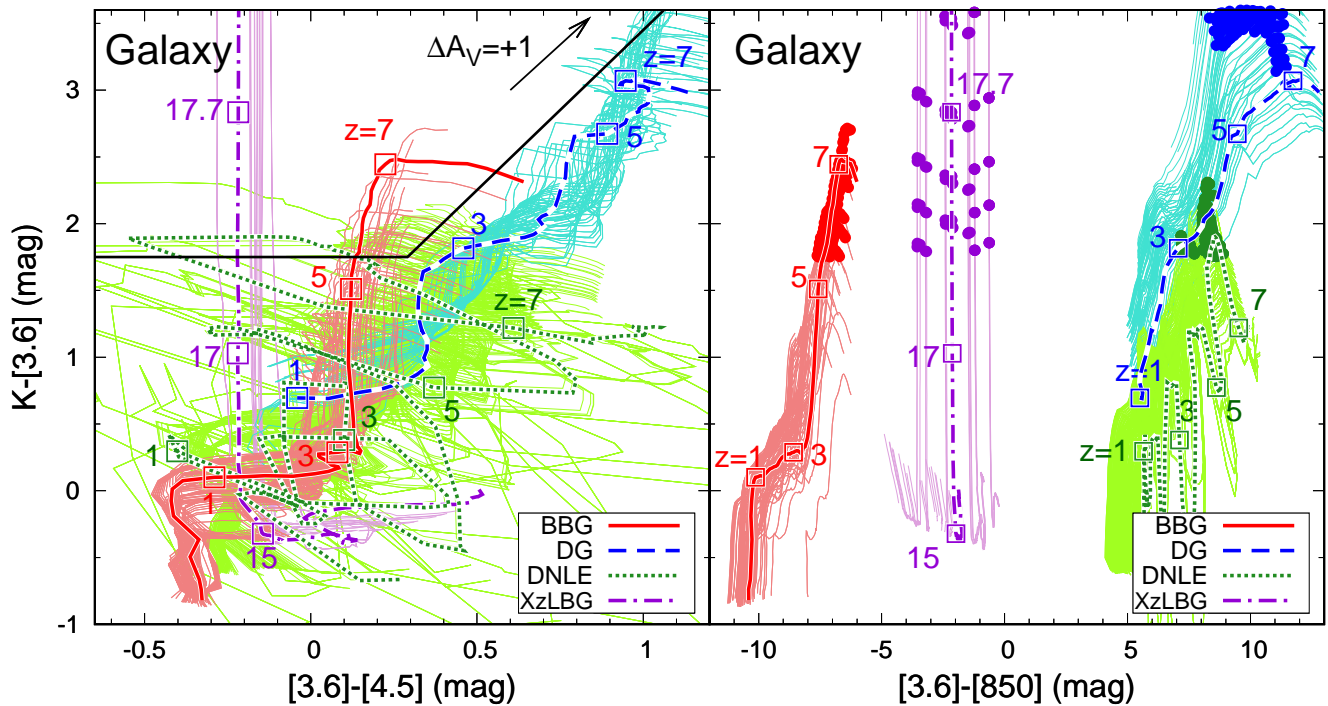}
\end{center}
\caption{The Star+Nebular+Dust model galaxy tracks from $z = 0$ to $8$ (or redshift when the age of the Universe is equal to the model age) on two-color diagrams of $K - [3.6]$ versus $[3.6] - [4.5]$ (left panel) and $K - [3.6]$ versus $[3.6] - [850]$ (right panel). BBGs, DGs, DNLEs, and XzLBGs are shown by the red, blue, green, and violet curves, respectively. For display purposes, we limited the dust attenuation to $A_V = 0$ (BBGs and XzLBGs), $4$ (DGs), and $2$ (DNLEs). The black arrow indicates the dust reddening effect in the case of $\Delta A_V = +1$. The thick curves are characteristic examples for each type of model galaxies shown in Figure~\ref{fig:modelspec}, where some redshifts are emphasized by squares. In the left panel, the BBG color criteria described in Equations~(\ref{eq:BBGcriteria1}) and (\ref{eq:BBGcriteria2}) are shown by the solid black line. In the right panel, the models satisfying the BBG color criteria are shown by the filled circles. \label{fig:TempColor}}
\end{figure*}

\begin{figure*}[]
\begin{center}
\includegraphics[width=1.0\linewidth, angle=0]{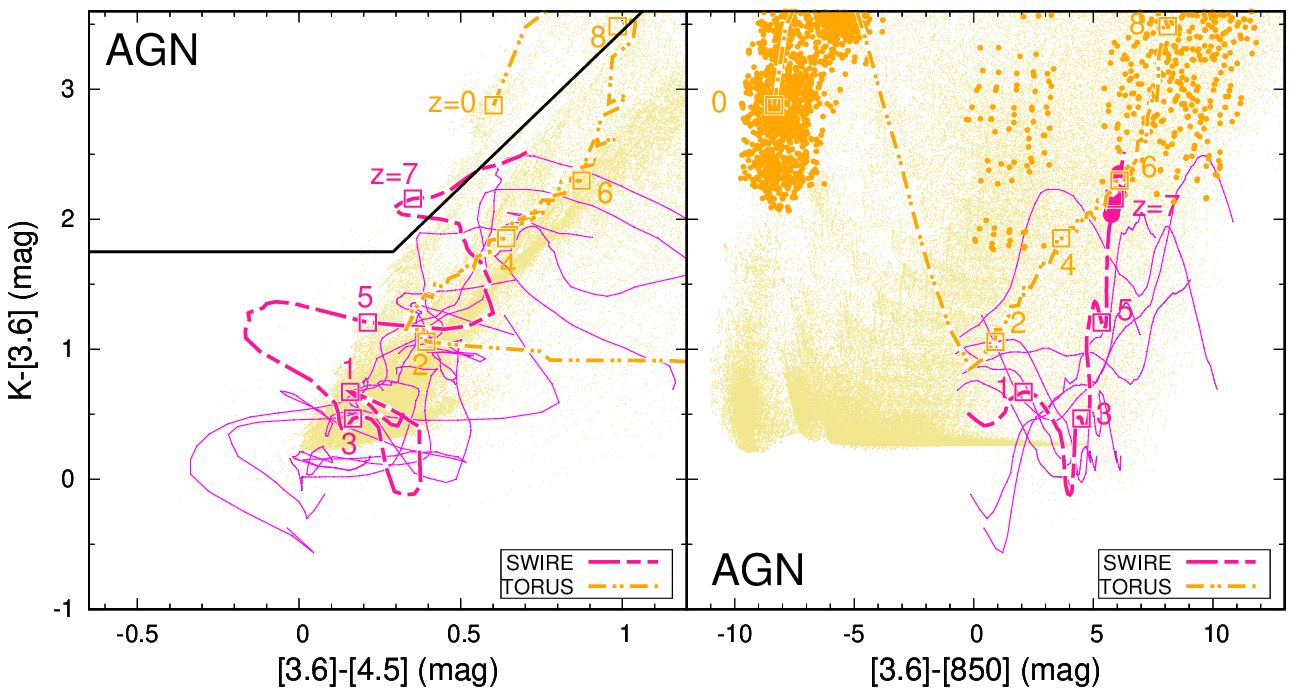}
\end{center}
\caption{Same as Figure~\ref{fig:TempColor} but for the AGN templates from the SWIRE library (magenta thin curves; \citealt{Polletta+07}) and the theoretical TORUS models (yellow dots; \citealt{Fritz+06,Feltre+12}). For the TORUS template colors, we only show those at limited redshifts ($z = 0, 1, 2, 3, 4, 5, 6, 7$, and $8$) for display purposes. The thick curves are characteristic examples: the SWIRE QSO2 template and one of the type-2 TORUS templates. \label{fig:TempColor+a}}
\end{figure*}

% Star+Nebular+Dust models with red K-[3.6] color 
We investigated colors of the Star+Nebular+Dust galaxy models of $0 < z < 8$ and $0 \leq A_V \leq 6$. We adopted the same setting for $T_{\rm age}$ and $Z$ as described in \S\ref{sec:modelgal}. Here, we only considered the exponentially declining SFH with $0.01\,{\rm Gyr} \leq \tau_{\rm SFH} \leq 10\,{\rm Gyr}$ to devise the BBG selection criteria. There are three types of galaxies that show extremely red $K - [3.6]$ colors: (1) passive galaxies dominated by old stars (BBGs) at $z$ $\ga 5$, (2) dusty galaxies (DGs) at $z > 1$, and (3) extremely young dusty star-forming galaxies with strong nebular emission lines (dusty nebular line emitters; DNLEs) at $z > 4$. Example spectra of these models are shown in Figure~\ref{fig:modelspec}. The Balmer break and dust attenuation make red $K - [3.6]$ colors for BBGs and DGs, respectively. For DNLEs, strong emission lines such as H$\alpha$ at $z \sim 4.5$ and [O~{\sc iii}]\,5007 at $z \sim 6.5$ boost the [3.6]-band flux. 
% (ここじゃないどこかに移動させる？)They are well separated in the space of the dust attenuation and the age normalized by the star-forming timescale ($\tau_{\rm SFH}$). The BBGs have $T_{\rm age} > 2 \times \tau_{\rm SFH} + 0.2\,{\rm Gyr}$ and $A_V \sim 0$, the DGs have $T_{\rm age} > \tau_{\rm SFH}$ and $A_V  > 2$, and the DNLEs have $T_{\rm age} < \tau_{\rm SFH}$ and $A_V  > 2$.

% Star+Nebular+Dust in K-[3.6] vs [3.6]-[4.5]; color selection criteria
The left panel of Figure~\ref{fig:TempColor} shows the Star+Nebular+Dust model tracks on the $K - [3.6]$ versus $[3.6] - [4.5]$ two-color diagram, from which we define the BBG color criteria as 
\begin{eqnarray}
K - [3.6] > 1.75, \label{eq:BBGcriteria1}\\
K - [3.6] > 2.4 ([3.6] - [4.5]) + 1.05. \label{eq:BBGcriteria2}
\end{eqnarray}
These color criteria are a slightly modified version of those in a previous study \citep{Mawatari+16b} to select BBGs with $T_{\rm age} \ga 0.3\,{\rm Gyr}$. 

% Comparison with rest-UVJ selection
At $z < 4$, a rest-frame $U - V$ and $V - J$ color selection method (rest-$UVJ$ selection; \citealt{Williams+09}) is often used to select BBGs. Based on the model template colors, we found that our BBG selection samples $\sim 0.1\,{\rm Gyr}$ younger galaxies than the rest-$UVJ$ selection. We neglect this small difference between the selection methods when comparing our results with other studies (\S\ref{sec:discussion}). 
We also confirmed that stars in the Milky Way do not satisfy the color criteria \citep{Mawatari+16b}. 

% Star+Nebular+Dust in [3.6]-[850] 
As can be seen in the left panel of Figure~\ref{fig:TempColor}, DGs and DNLEs can contaminate our BBG selection criteria. DGs and DNLEs can be removed by dust emission in FIR. 
%and young stellar emission around $\lambda \sim 1\,\mu$m (i.e., rest-frame UV), respectively. 
BBGs are dust-poor and much bluer in ${\rm NIR} - {\rm FIR}$ color than DGs and DNLEs. In the right panel of Figure~\ref{fig:TempColor}, we show $[3.6] - [850]$ colors (i.e., NIR $-$ FIR colors) of the Star+Nebular+Dust model galaxies\footnote{Here, we fix $[3.6] = 24$\,mag, which is almost the same as observed magnitudes for our final BBG sample (\S\ref{sec:BBGselect} and Table~\ref{tb:obsprop}), to calculate the model's $L_{\rm IR}$ and then $850\,\mu$m magnitude.}. The $[3.6] - [850]$ colors of BBGs are clearly different from others, as expected. 
%On the other hand, no notable difference in the $Y - [3.6]$ colors can be seen. The $Y - [3.6]$ colors of DNLEs at $4 \lesssim z \lesssim 7$ are intrinsically blue but observed to be as red as BBGs due to dust reddening.

% Interesting possibility: LBGs at z > 15
Additionally, we note an interesting possibility that our BBG criteria can identify Lyman break galaxies (LBGs) at extremely high-redshift (extremely high-$z$ LBGs; XzLBGs). In Figures~\ref{fig:TempColor}, the XzLBG color tracks at $z > 10$ are superposed based on a Stellar+Nebular+Dust model with $Z = 0.004$, $T_{\rm age} < \tau_{\rm SFH}$ and $A_V = 0$. The XzLBGs at $17.5 \lesssim z \lesssim 30$ satisfy the BBG criteria. An example spectrum of the XzLBG models is also shown in Figure~\ref{fig:modelspec}. 

% AGN in 2 CC diagrams
We also investigated colors of the AGN templates. We found that our BBG color criteria can be satisfied by the SWIRE QSO2 template at $z \sim 7$ as well as by some type-2 TORUS templates at $z \lesssim 1$ and $z \sim 8$ (the left panel of Figure~\ref{fig:TempColor+a}). Their spectra are also shown in Figure~\ref{fig:modelspec}. In the SWIRE QSO2 case, the red $K - [3.6]$ color can be achieved by a combination of heavily obscured continuum and the broad [O~{\sc iii}]\,5007 emission line around $3.6\,\mu$m. In the TORUS model case, torus continuum emission alone can mimic the BBG-like color. Among these AGN contaminations, the SWIRE QSO2 type objects can be removed by their bright FIR emission (the right panel of Figure~\ref{fig:TempColor+a}). In contrast, some of the TORUS contaminations have blue $[3.6] - [850]$ color, which makes it hard to distinguish them from the BBGs at $z \gtrsim 5$. However, such AGNs with very little contribution from the host galaxies to the whole SED would be extreme and rare, as we discuss later (\S\ref{sec:disc_z0AGN}).

\section{Selection of BBG Candidates}\label{sec:BBGselect}

In this section, we present the selection procedure of BBG candidate galaxies from the multiwavelength data in the COSMOS field. First, we select BBG candidates on the $K-[3.6]$ versus $[3.6] - [4.5]$ two-color diagram (\S\ref{sec:CCsel}). Then, we narrow down the candidates to six that are not detected in X-ray, FIR, and radio bands as well as in optical bands (\S\ref{sec:StrictSelection}).

\subsection{NIR Color Selection}\label{sec:CCsel}

Source extraction was performed on the SPLASH [3.6]-band image using SExtractor \citep{BertinArnouts96} version 2.5.0. We masked areas around objects brighter than 20 mag in [3.6] to remove faint objects whose photometry was affected by the bright objects. The masked area of each bright object was defined by the isophotal level at twice the sky fluctuation in the [3.6]-band image. Avoiding the masked region, the effective area was $0.41$\,deg$^2$. We focused on isolated sources within a $3$\,arcsec radius not only in the [3.6]-band image but also in the $K$- and [4.5]-band images. Namely, we selected sources that do not have any nearby objects brighter than 10\,\% of its flux density in all of the three bands within the circular area. We found $\sim 37,000$ such isolated objects down to $[3.6] \approx 24.1$\,mag ($4\,\sigma$) and called them the parent sample. 

\begin{figure}[]
\begin{center}
\includegraphics[width=1.0\linewidth, angle=0]{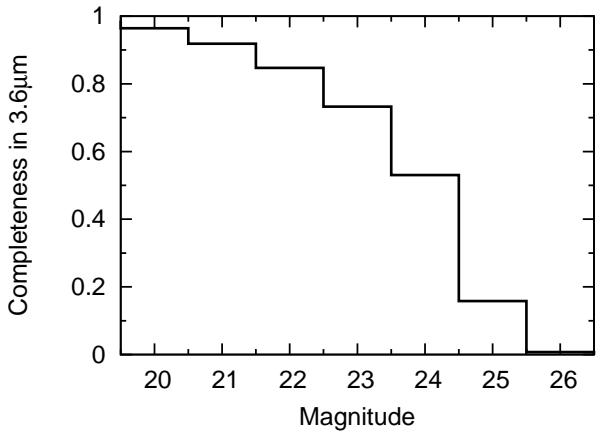}
\end{center}
\caption{Completeness of source extraction in the [3.6]-band image. This is evaluated by extracting artificial objects in the same manner as the parent sample selection (see the text). \label{fig:completeness}}
\end{figure}

We estimated the completeness of our source extraction by detecting artificial sources randomly embedded in the [3.6]-band image. Following \citet{Barmby+08,Ashby+13}, we considered that an artificial object is recovered if the object is detected within 1\,arcsec from the input position and its measured flux density is within a 50\,\% difference from the priori flux density. We further adopted the isolation criterion to match the parent sample construction. The resultant completeness as a function of the input artificial objects' magnitude is shown in Figure~\ref{fig:completeness}. We could detect objects brighter than $\sim 24$\,mag with completeness higher than 50\,\%. 
%The low completeness at faint flux level in the [3.6] image does not affect the following analysis so match, because detectability for the BBGs is limited mainly by the depth in the $K$-band rather than in the [3.6]. 

In photometry, we used the task ``PHOT'' of Image Reduction and Analysis Facility (IRAF), where aperture diameters were set to $2 \times$ the FWHM of PSFs in every band image. The photometric aperture is centered at the detection position in each band. If no object was detected in the $K$- or [4.5]-band images within $1\,$arcsec from the [3.6]-band detected position, the photometric aperture was forced to be centered at the [3.6]-band position. The aperture magnitudes were corrected to the total magnitudes using aperture correction factors estimated for point sources. For the photometric uncertainty, we measured $1\,\sigma$ of the distribution of random aperture photometry in each image and applied the same aperture correction. We neglected the Galactic extinction for the $K$, [3.6], and [4.5] photometry because it is very small ($<0.01$\,mag; \citealt{Schlegel+98} with $R_V = 3.1$). 

We applied the BBG color criteria (Equations~\ref{eq:BBGcriteria1} and \ref{eq:BBGcriteria2}) to the parent sample down to $\approx 24.1$\,mag corresponding to the $4\,\sigma$ limiting magnitude in the [3.6]-band. In the case of non-detection ($< 2\,\sigma$) in the $K$- and [4.5]-bands, we put the lower and upper limits on their $K - [3.6]$ and $[3.6] - [4.5]$ colors with the $2\,\sigma$ limiting magnitudes, respectively. We identified 23 objects satisfying the BBG color criteria, which are referred to as the color-selected sample in the following sections.

\subsection{Multi-band Selection}\label{sec:StrictSelection}

We constructed a multi-band photometry catalog for the color-selected sample. For the photometry at wavelengths between $0.4\,\mu$m and $10\,\mu$m, we measured the total magnitudes ourselves in the same manner as adopted for the $K$-, [3.6]-, and [4.5]-band images. Here, we used $2 \times {\rm PSF}$ apertures for all bands but the $F814W$-band. For the $F814W$-band photometry, larger ($0.6$\,arcsec diameter) apertures were used to avoid flux loss by a possible spatial offset between the $HST$ image and the [3.6]-band image, the latter of which has much coarser resolution. The measured magnitudes were corrected for Galactic extinction with $A_{F814W} = 0.03$, $A_{g} = 0.07$, $A_{r} = 0.05$, $A_{i} = 0.04$, $A_{z} = 0.03$, $A_{y} = 0.03$, $A_{Y} = 0.02$, $A_{J} = 0.02$, $A_{H} = 0.01$, and $A_{\rm [5.8]} = A_{\rm [8.0]} = 0$, which were estimated for the center of the COSMOS field based on \citet{Schlegel+98}. For wavelengths longer than $10\,\mu$m and X-ray, we used the publicly available catalogs constructed by the individual survey teams (\S\ref{sec:data}). 

\begin{figure*}[]
\begin{center}
\includegraphics[width=1.0\linewidth, angle=0]{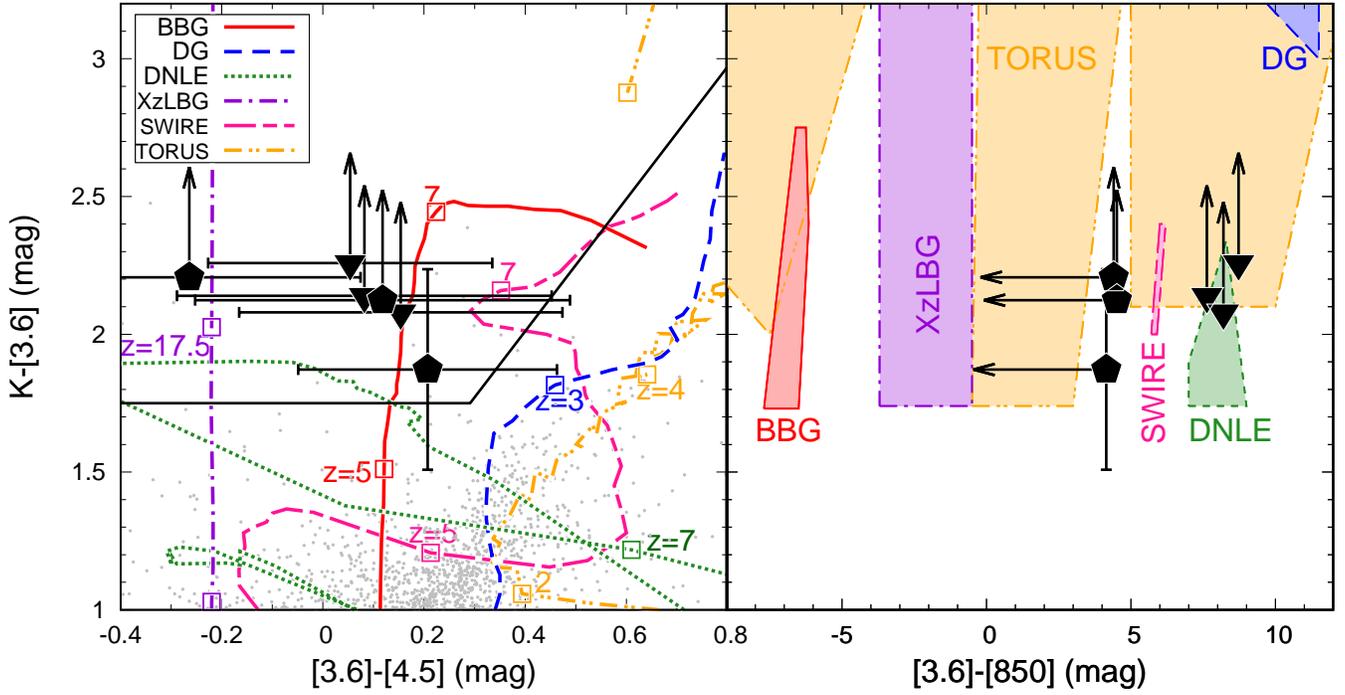}
\end{center}
\caption{Color distributions of the BBG candidates with and without ALMA detections are shown with filled triangles and pentagons, respectively, in the $K - [3.6]$ versus $[3.6] - [4.5]$ (left) and $K - [3.6]$ versus $[3.6] - [850]$ (right) diagrams. Here, we used the ALMA Band\,7 data for the $850\,\mu$m photometry (\S\ref{sec:ALMA}). The arrows indicate $2\,\sigma$ limits for non-detections in $K$ or ALMA Band\,7. In the left panel, the characteristic template tracks are superposed by thick curves, which are the same as those in Figures~\ref{fig:TempColor} and \ref{fig:TempColor+a}. We also show $K - [3.6]$ and $[3.6] - [4.5]$ colors of all objects in the parent sample (grey dots). The BBG color selection boundary is shown by the black solid line. In the right panel, the shaded polygons correspond to the areas occupied by the Star+Nebular+Dust and AGN templates satisfying the BBG color criteria (Equations~\ref{eq:BBGcriteria1} and \ref{eq:BBGcriteria2}). \label{fig:pTRG_colors}}
\end{figure*}

\begin{deluxetable*}{lcccccc}
\tablecaption{Observed properties of the BBG candidates \label{tb:obsprop}}
\tablehead{
\colhead{Name}&\colhead{R.A.}&\colhead{Dec.}&\colhead{$K$}&\colhead{[3.6]}&\colhead{[4.5]}&\colhead{ALMA Band\,7}\\
\colhead{}&\colhead{(degree)}&\colhead{(degree)}&\colhead{(mag)}&\colhead{(mag)}&\colhead{(mag)}&\colhead{(mag)}
}
%\colnumbers
\startdata
\multicolumn{7}{c}{\bf{\normalsize Sample without ALMA detections}} \\
SPLASH\_COSMOS\_z6BBG\_09 & 149.680749 & 2.062202 & 25.61$\pm$0.31 & 23.74$\pm$0.19 & 23.54$\pm$0.17 & $>$ 19.60 \\
SPLASH\_COSMOS\_z6BBG\_22 & 150.071625 & 2.645838 & $>$ 26.2 & 23.99$\pm$0.23 & 24.25$\pm$0.25 & $>$ 19.59 \\
SPLASH\_COSMOS\_z6BBG\_29 & 149.724422 & 1.757402 & $>$ 26.2 & 24.10$\pm$0.27 & 23.99$\pm$0.26 & $>$ 19.59 \\
\hline
\multicolumn{7}{c}{\bf{\normalsize Sample with ALMA detections}} \\
SPLASH\_COSMOS\_z6BBG\_19 & 150.074593 & 2.045192 & $>$ 26.2 & 23.94$\pm$0.22 & 23.88$\pm$0.18 & 15.21$\pm$0.03 \\
SPLASH\_COSMOS\_z6BBG\_27 & 149.825170 & 2.084366 & $>$ 26.2 & 24.09$\pm$0.26 & 24.01$\pm$0.26 & 16.46$\pm$0.09 \\
SPLASH\_COSMOS\_z6BBG\_30 & 150.220922 & 2.607786 & $>$ 26.2 & 24.11$\pm$0.26 & 23.96$\pm$0.19 & 15.91$\pm$0.07 \\
\enddata
\tablecomments{The $2\,\sigma$ limiting magnitudes are shown for the fainter objects. For the ALMA Band\,7, we assume point sources to put the upper constraints.}
\end{deluxetable*}

As shown in Figure~\ref{fig:TempColor+a}, an additional criterion of $[3.6] - [850] < -5$ should remove DGs, DNLEs and type-2 AGNs except for objects with spectra similar to the $z \sim 0$ TORUS model. We practically adopted non-detection in all available FIR data whose depths were much shallower than those of the SPLASH [3.6]-band (Table~\ref{tb:data_sum}). X-ray and radio data were also useful to remove AGNs. DNLEs may be detected in the optical bands if the Ly$\alpha$ emission line is strong enough to boost the broad band flux. 

No object in the color-selected sample was matched with any source in the X-ray and FIR catalogs. One object matched with a source in the VLA 3\,GHz catalog \citep{Smolcic+17}, which was removed from our sample. 
We discarded 15 objects detected in some of the optical bands (shorter than $Y$) with a significance more than $2\,\sigma$. We further excluded an object because its [3.6]-band photometry was obviously affected by a nearby extended galaxy. The remaining six objects were recognized as BBG candidates at $5 \lesssim z \lesssim 8$. Their coordinates and photometry are shown in Table~\ref{tb:obsprop}. Hereafter, their names, SPLASH\_COSMOS\_z6BBG\_XX, are simplified as ``BBG\_XX''. Their sky and color distributions\footnote{The $K$-band weight map released by the UltraVISTA team reveals homogeneous local sky variance in the UVISTA UD stripes 2 and 3. Only BBG\_30 among the six candidates lies at the edge of the $K$-band image where local sky variance is larger than $1.2 \times$ the average in the UD stripes. Even if we conservatively adopt a $0.2$\,mag shallower limiting magnitude in the $K$-band than the representative (Table~\ref{tb:data_sum}), all of the six candidates satisfy the $K - [3.6]$ color criterion (Figure~\ref{fig:pTRG_colors}).} are shown in Figures~\ref{fig:coverage} and \ref{fig:pTRG_colors}, respectively. Figure~\ref{fig:multiIM_ALMABBG} shows the multi-band images of the six BBG candidates, where we select representative bands among many non-detection images. 

At $0.4\,\mu$m$ < \lambda < 1\,\mu$m, we also checked publicly available catalogs from the $HST$-COSMOS and HSC-SSP. Among the six candidates, only BBG\_29 is in the HSC-SSP catalog \citep{Aihara+19}. Its $i$-band magnitude in the catalog is $\sim 27$\,mag, corresponding to $S/N \sim 5$. By close inspection, we found that the catalog magnitude of BBG\_29 was actually overestimated because of the locally enhanced background sky fluctuation. Flux measurements of the marginally detected objects are sensitive to photometric parameters such as an aperture size and a width of annulus for estimation of the sky level. We again measured the flux of all six candidates in all HSC broad-bands adopting various combinations of the aperture sizes ($=$ 1--2 $\times {\rm PSF}$) and the sky annuli (5--21\,arcsec). The resultant fluxes were very faint with typically $S/N < 2$ for any candidates in any bands. Exceptions were found for the $i$- and $z$-bands of BBG\_22 as well as the $i$-band of BBG\_29, where $S/N$ reached as high as $\sim 3$ by few combinations of the aperture sizes and sky annuli. Even with the brightest measurements in the $i$- and $z$-bands ($i$-band) for BBG\_22 (BBG\_29), our results based on the SED analysis remain unchanged (\S\ref{sec:SEDfit}).

\section{Follow-up ALMA Band\,7 observations}\label{sec:ALMA}

%\begin{figure*}[httb]
\begin{figure*}[]
\begin{center}
\includegraphics[width=1.0\linewidth, angle=0]{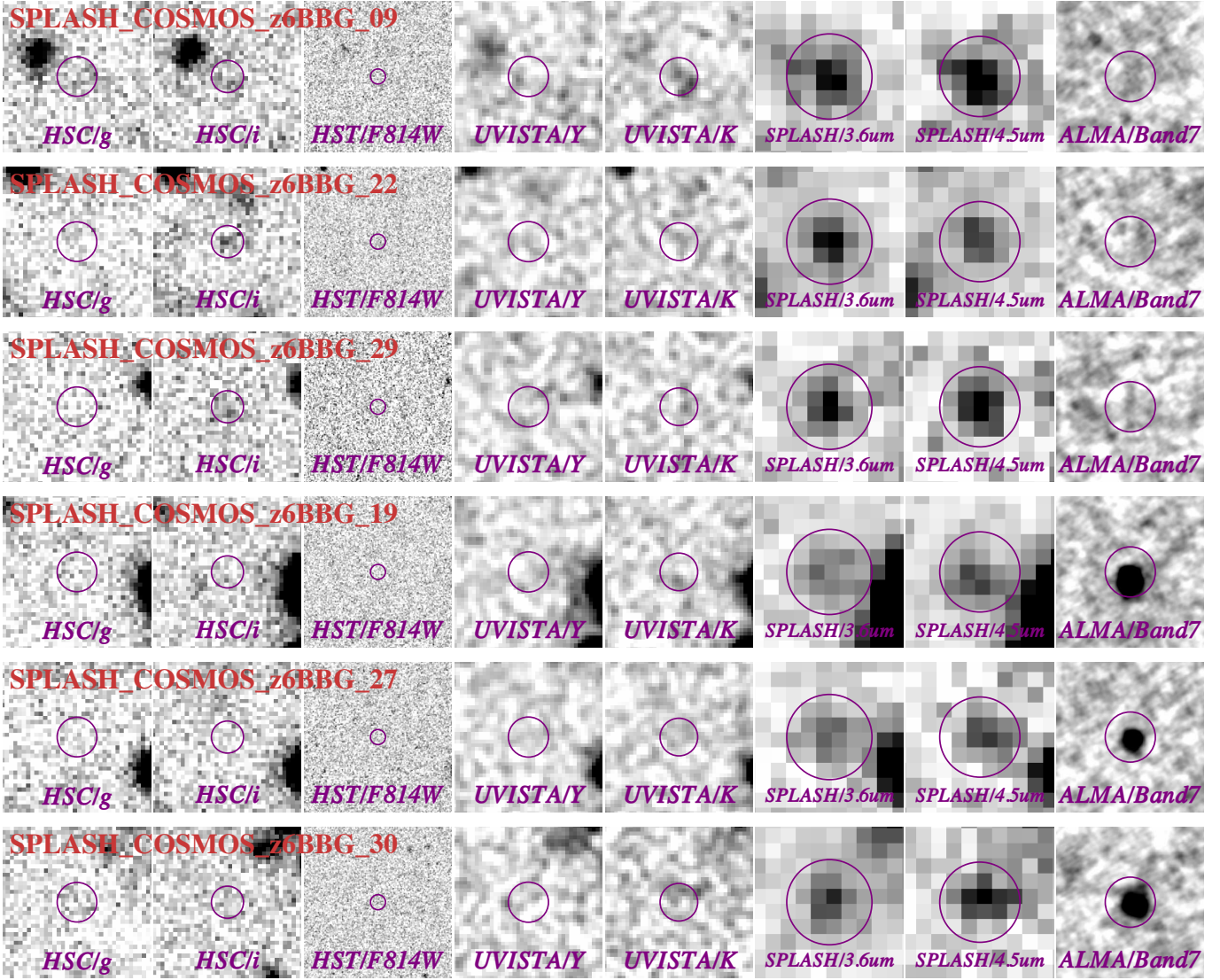}
\end{center}
\caption{Postage images of the six BBG candidates in $g$, $i$, $F814W$, $Y$, $K$, [3.6], [4.5], and ALMA Band\,7. The top three objects are not detected in ALMA Band~7, whereas the remaining three objects are detected. The panel size is always $6'' \times 6''$. The circle superposed on each panel shows the photometric aperture with diameters of $0\arcsec.6$ for $F814W$, $2\arcsec$ for ALMA Band\,7, and $2 \times$ FWHM of the PSF for the other bands. \label{fig:multiIM_ALMABBG}}
\end{figure*}

The $Herschel$ and JCMT/SCUBA-2 data in the COSMOS field are the deepest among the existing wide-field FIR images and allow us to distinguish BBGs from dusty contaminants such as DGs and DNLEs marginally. For conclusive discrimination, however, the FIR limit had to be deepened. Therefore, we conducted ALMA follow-up observations for the BBG candidates.
%While the $Herschel$ and JCMT/SCUBA-2 data used in the previous section are the deepest ones among existing FIR data from surveys covering the entire COSMOS field, they are still not enough to discriminate the BBGs at $z \gtrsim 5$. For example, the current SCUBA-2depth limit (Table~\ref{tb:data_sum}) corresponds to a upper limit on the infrared color of $[3.6] - [850] \lesssim 9$ for the BBG candidates with $[3.6] \sim 24$, which is not enough to distinguish the BBGs from dusty contamination such as the DGs and DNLEs (Figure~\ref{fig:TempColor+a}). 

The ALMA Band\,7 observations (ID 2017.1.01259.S, P.I.: K. Mawatari) were performed in Cycle 5. We observed the six BBG candidates and additional six filler objects. The observations were performed in April, May, August, September, and October 2018 under the antenna configurations of C43-2, C43-3, C43-4, and C43-5. The total on-source integration time was $38.8$\,minutes for each target. Four spectral windows (SPWs) with a total band width of $7.5$\,GHz were set at the central frequency of $336.5$, $338.4$, $348.5$, and $350.5$\,GHz. The corresponding wavelength coverages were $\lambda = 853$ -- $863$\,$\mu$m and $\lambda = 883$ -- $893$\,$\mu$m. The spectral resolution was set to $15.6$\,MHz in the time division mode (TDM), which is enough to measure the continuum. The following six QSOs were used for calibrations: J1058+0133, J0948+0022, J1037-2934, and J0854+2006 for atmospheric calibration; J1058+0133, J0948+0022, J0942-0759, J1037-2934, and J0854+2006 for water vapour radiometer (WVR) calibration; J1058+0133, J1037-2934, J0942-0759, and J0854+2006 for pointing calibration; J1058+0133, J1037-2934, and J0854+2006 for bandpass and flux calibration; and J0948+0022 for phase calibration. According to ALMA proposer's Guide, the flux calibration uncertainty is expected to be less than $10$\,\% in Band\,7. 

The data reduction and calibration were performed using the Common Astronomy Software Application (CASA) pipeline version 5.4.0. We collapsed all channels to produce a dust continuum image using the CASA task, {\tt\string CLEAN}, with the natural weighting. 
% synthesized beam size and PA come from the Takuya's measurement file.
The resulting synthesized beam size in FWHM was $0\arcsec.48 \times 0\arcsec.42$ with a position angle PA $\approx -78\arcdeg$. We achieved $1\,\sigma$ RMS level of $\sim 30\,\mu$Jy\,beam$^{-1}$ for all the target objects. Photometry for the 12 targets was performed on the dust continuum images using CASA task {\tt\string imfit} that fits the observed data within $2\arcsec$ diameter apertures centered at the [3.6]-band detected positions with 2D Gaussian light profiles. 

Three among the six BBG candidates (eight out of the 12 targets in total) were detected in the continuum images. We hereafter focus on the BBG candidates, whose postage ALMA images are shown in Figure~\ref{fig:multiIM_ALMABBG}. Their ALMA Band\,7 flux density measurements are summarized in Table~\ref{tb:obsprop}. For the three BBG candidates not detected in ALMA Band~7, we obtained the flux density upper limits assuming point-like sources. The constraint on the FIR photometry became about two orders of magnitude deeper than the $Herschel$ and SCUBA-2 data. Using the ALMA Band\,7 continuum flux density, we plotted the $[3.6] - [850]$ colors of the six BBG candidates in the right panel of Figure~\ref{fig:pTRG_colors}. The colors of the three BBG candidates without any ALMA detection are hard to be explained by the DG and DNLE models or SWIRE AGN templates. We therefore conclude that they are the most likely BBG candidates at $5 \lesssim z \lesssim 8$, although the contamination from dusty tori of type-2 AGNs or XzLBGs cannot be completely ruled out.

\section{SED Fitting}\label{sec:SEDfit}

\subsection{Fitting Method}\label{sec:fitmethod} % basically present style for the method description

\begin{table*}[]
%\rotate
\begin{center}
\caption{SED fitting parameters in the three template groups\label{tb:SEDfitset}}
\begin{tabular}{cllll}
\hline
\hline
Group name & \multicolumn{1}{c}{Galaxy} & \multicolumn{2}{c}{AGN} & \multicolumn{1}{c}{XzLBG} \\
\cmidrule(lr){2-2} 
\cmidrule(lr){3-4} 
\cmidrule(lr){5-5} 
Template type\tablenotemark{a} & Star+Nebular+Dust & SWIRE AGN & TORUS & Star+Nebular+Dust \\ 
\hline
Number of templates & 2,818,260 & 7,560 & 960,000 & 12,030 \\
\hline
SFH & Exp-declining/rising & --- & --- & Constant-SFR \\
 & ($\tau_{\rm SFH} = \pm0.03$, $\pm0.06$, $\pm0.1$, &  &  &  \\
 &  $\pm0.3$, $\pm0.6$, $\pm1$, and $\pm10$\,Gyr), & &  &  \\
 & Constant-SFR &  & &  \\
\hline
Metallicity ($Z$) & 0.0001, 0.004, and 0.02 & --- & --- & 0.004 \\
\hline
Age ($T_{\rm age}$) [Gyr] & 0.001 -- age of the Universe & --- & --- & 0.001 -- age of the Universe \\ 
\hline
Redshift\tablenotemark{b} & 0.1 -- 7.9 & 0.1 -- 7.9 & 0.1 -- 7.9 & 10.1 -- 29.9 \\
\hline
$A_V$\tablenotemark{c} [mag] & 0 -- 10 & $-2$ -- $+2$ & 0 & 0 -- 0.5 \\
\hline
\end{tabular}
\end{center}
\tablenotetext{a}{Three types of the spectral templates are used: ``Star+Nebular+Dust'' is our galaxy spectral model including stellar, nebular, and dust emissions; ``SWIRE AGN'' denotes the empirical AGN templates from the SWIRE library \citep{Polletta+07}; ``TORUS'' is the theoretical dust torus model \citep{Fritz+06,Feltre+12}. }
\tablenotetext{b}{Redshift steps are $\Delta z = 0.2$.}
\tablenotetext{c}{Dust attenuation step is $\Delta A_V = 0.2$ except for XzLBG group where $\Delta A_V = 0.1$ is adopted. For the Star+Nebular+Dust templates, dust attenuation $A_V$ is limited to $A_V < \rm{max} (4 \times SFR^{0.3}, 3.5)$ (see Appendix~\ref{sec:Av2SFR}).}
\end{table*}

We performed SED fitting analyses of the multi-band photometric data at wavelengths between $0.4\,\mu$m and $1000\,\mu$m (from HSC/$g$ to ALMA/Band\,7) for the six BBG candidates presented in \S\ref{sec:BBGselect}. We mainly discuss the three candidates without ALMA detection in the following sections. Results for the other three candidates with ALMA detection are shown in Appendix~\ref{sec:Fit4ALMAcontami}. The fitting code used in this study is our original SED analysis package (``PANHIT'') that is publicly available from our website\footnote{http://www.icrr.u-tokyo.ac.jp/$\sim$mawatari/PANHIT/PANHIT.html}. 
% chi2 def
We followed a $\chi^2$ minimization algorithm for data including upper limits proposed by \citet{Sawicki+12} but modified the formula slightly. Our definition of the $\chi^2$ is as follows:
\begin{eqnarray}
\chi^2 & = & \sum_{i} \left(\frac{f_{{\rm obs},i} - s f_{{\rm temp},i}}{\sigma_{{\rm obs},i}}\right)^2 -2 \sum_{j}   \nonumber \\
       &  &  \ln \left[ \frac{1}{\sqrt{2\pi}\sigma_{{\rm obs},j}} \int^{f_{{\rm lim},j}}_{-\infty} \exp \left\{ -\frac{1}{2} \left( \frac{f - s f_{{\rm temp},j}}{\sigma_{{\rm obs},j}} \right)^2 \right\} df \right] , \nonumber \\ \label{eq:chi2_SEDfit}
\end{eqnarray} 
where $f_{\rm obs}$, $\sigma_{\rm obs}$, and $f_{\rm temp}$ are the observed flux density, its uncertainty, and the template flux density, respectively. In Equation~(\ref{eq:chi2_SEDfit}), the indices $i$ and $j$ in the summations of the first and second terms in the right-hand side correspond to the detection and non-detection bands, respectively. 
% det/nondet
Here, we regard the flux density brighter (fainter) than 2$\sigma$ limit as a detection (non-detection) for the bands at $\lambda < 10$\,$\mu$m. At longer wavelengths, the BBG candidates are not detected in all bands (\S\ref{sec:StrictSelection}) except for the three candidates in the ALMA Band\,7 (\S\ref{sec:ALMA}). Following treatment in \citet{Sawicki+12}, the upper limit of the integral in the non-detection band term, $f_{\rm lim}$, is set to the $1\,\sigma$ limiting flux density. 
%The non-detection band term approaches a small but non-zero value when the template flux density is much smaller than the flux limit ($s f_{{\rm temp},j} \ll f_{{\rm lim},j}$), which is different from the detection band term that becomes zero when the template flux density is equal to the observed measurement ($s f_{{\rm temp},i} = f_{{\rm obs},i}$). 
% scaling
The scaling factor, $s$, is estimated analytically using only the detection bands: 
\begin{eqnarray}
s = \sum_{i} \frac{f_{{\rm obs},i} f_{{\rm temp},i}}{\sigma_{{\rm obs},i}^2} \bigg/ \sum_{i} \frac{f_{{\rm temp},i}^2}{\sigma_{{\rm obs},i}^2} .\label{eq:s_SEDfit}
\end{eqnarray}
% Mention about flux upper limit? 

Three groups of templates were prepared for SED fitting. Fitting parameter ranges in each template group are summarized in Table~\ref{tb:SEDfitset}. The first group, called the Galaxy group, consists of the Star+Nebular+Dust templates (\S\ref{sec:modelgal}) at $0 < z < 8$ with a wide variety of physical parameters. For the exponentially declining/rising SFH in the Star+Nebular+Dust models, we restricted the e-folding timescale to $\tau_{\rm SFH} \geq 30$\,Myr. The lower limit of $\tau_{\rm SFH}$ was determined to make it comparable to a free-fall time of a spherically symmetric system virialized at $z \sim 20$ \citep{Mo+10}. The $\tau_{\rm SFH} = 30$\,Myr is much longer than the gas cooling time for a halo with gas mass less than $10^{10}\,M_\odot$ at $z > 10$ \citep{Mo+10}. The second template group, called the AGN group, consists of AGN templates (\S\ref{sec:tempAGN}). The empirical SWIRE AGN templates that are already reddened by dust are further reddened or dereddened by $-2 \leq A_V \leq +2$. We did not consider any dust reddening for the theoretical TORUS templates. The third template group, called the XzLBG group, was for possible XzLBG solutions. This third group consists of the Star+Nebular+Dust templates with a redshift range as high as $10 < z < 30$ and relatively simplified settings for the other model parameters (Table~\ref{tb:SEDfitset}). We performed SED fitting separately with the above three template groups because we cannot know which template type is physically probable for each BBG candidate in advance. 

For dust attenuation, while we mainly show the results with the Calzetti law \citep{Calzetti+00} in the following sections, we also adopted the Milky Way (MW) law \citep{Seaton79} and the Small Magellanic Cloud (SMC) law \citep{Prevot+84}. In our preliminary SED analyses, we occasionally obtained peculiar solutions with an almost zero SFR and extremely high dust attenuation. These solutions seem to be unlikely, and we define the forbidden region in the $A_V$--SFR plane as $A_V > \rm{max}(4 \times SFR^{0.3}, 3.5)$ (see Appendix~\ref{sec:Av2SFR} for details).

We adopted a Monte-Carlo (MC) technique to evaluate the reliability of the fitting solutions. We repeated the SED fitting procedures for randomly perturbed SEDs. The perturbation added to the observed flux density was realized by drawing a random number from a Gaussian distribution whose standard deviation is equal to the 1$\sigma$ uncertainty in each band. The distribution of the best-fit solutions in these MC realizations defines the probability of the fitting solutions as well as the confidence intervals around the solutions. To avoid confusion, hereafter, we refer to the template yielding the least $\chi^2$ in the fitting to the actual observed SED as the ``best-fit'' template and one derived from each MC realization as the ``MC-best'' template.

\subsection{BBG candidates without ALMA detection}\label{sec:Fit4ALMABBG}

Here, we present the SED analyses for the three BBG candidates not detected in the ALMA observations. Because the ALMA Band~7 upper limits are deeper than $Spitzer$/MIPS, $Herschel$, and JCMT/SCUBA-2 data, we used only the ALMA Band~7 data for the FIR range in the SED fitting. The resultant number of the bands is 15: $F814W, g, r, i, z, y, Y, J, H, K_s, [3.6], [4.5], [5.8], [8.0]$, and ALMA Band~7. In the following sections, we describe the SED fitting performed separately with the three template groups (\S\ref{sec:Fit4ALMABBG_fid}) and that performed with composite templates of the galaxy and AGN models (\S\ref{sec:Fit4ALMABBG_sup}). 

 %\begin{figure}[httb]
\begin{figure*}[]
\begin{center}
\includegraphics[width=1.0\linewidth, angle=0]{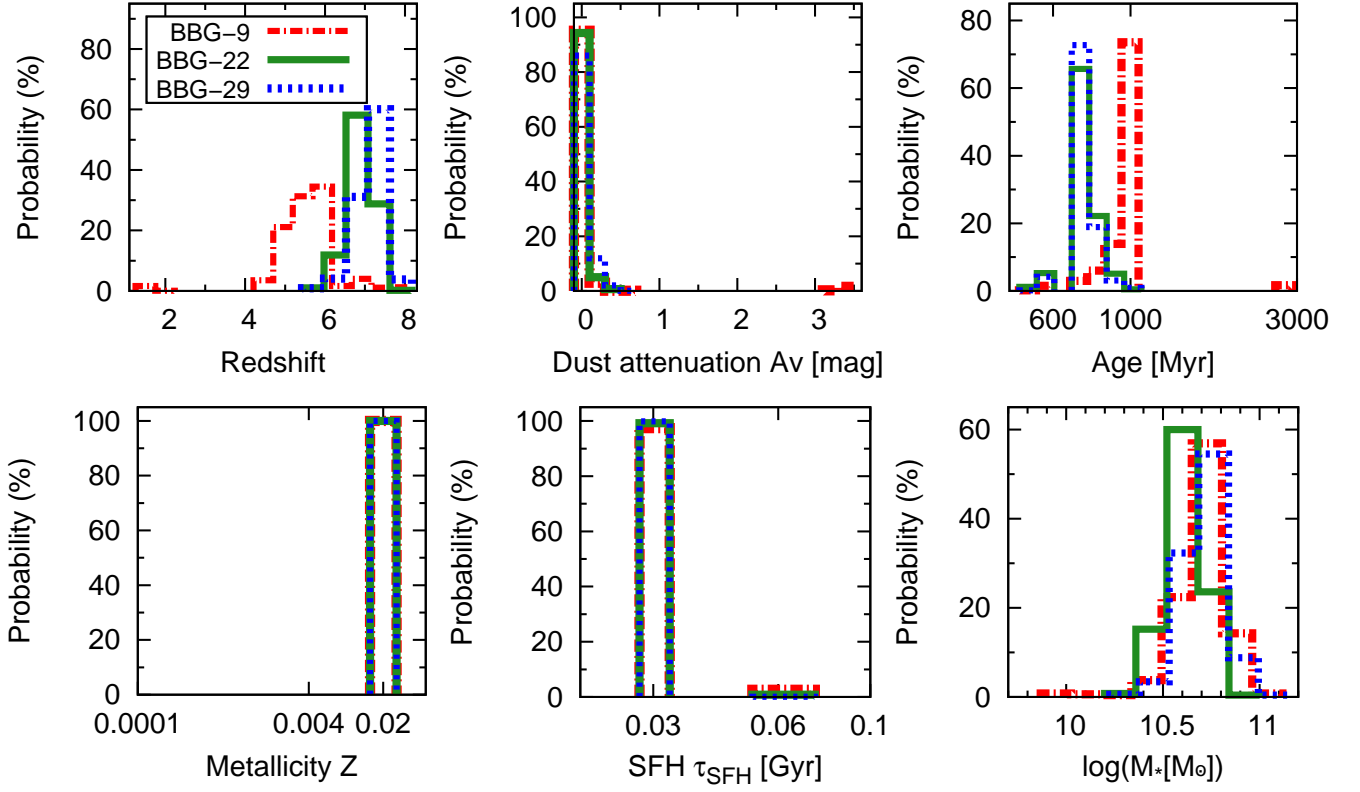}
\end{center}
\caption{Probability distributions of physical quantities for the BBG candidates without ALMA detection. These are derived from 1,000 MC realizations of the SED fitting with Galaxy group templates (see Table~\ref{tb:SEDfitset}). \label{fig:MChist}}
\end{figure*}

\begin{table*}
\caption{Physical properties of the best-fit BBG models} \label{tb:fitprop}
\begin{center}
\begin{tabular}{clll}
\hline
\hline
 & BBG\_9 & BBG\_22 & BBG\_29 \\
\hline
$\chi^2$ & 6.2 & 9.1 & 7.2 \\
Redshift ($z$) & $5.5^{+0.6}_{-0.6}$ & $6.9^{+0.3}_{-0.4}$ & $7.1^{+0.5}_{-0.5}$ \\
Dust attenuation (A$_{V}$) [mag] & $0.0^{+0.2}_{-0.0}$ & $0.0^{+0.2}_{-0.0}$ & $0.0^{+0.2}_{-0.0}$ \\
Age ($T_{\rm age}$) [Gyr] & $1.02^{+0.0}_{-0.18}$ & $0.72^{+0.14}_{-0.13}$ & $0.72^{+0.12}_{-0.13}$ \\
Metallicity ($Z$) & $0.02$\tablenotemark{a} & $0.02$\tablenotemark{a} & $0.02$\tablenotemark{a} \\
Star formation timescale ($\tau_{\rm SFH}$) [Gyr] & $0.03^{+0.02}_{-0.00}$ & $0.03^{+0.02}_{-0.00}$ & $0.03^{+0.02}_{-0.00}$ \\
Stellar mass ($M_*$) [$10^{10}\,M_\odot$] & $5.2^{+1.1}_{-1.3}$ & $4.1^{+0.9}_{-0.8}$  & $5.1^{+1.3}_{-1.0}$ \\
\hline
\end{tabular}
\end{center}
\tablecomments{The uncertainties are the 68\,\% ranges of the distributions of the MC-best models in the SED fitting with Galaxy group templates. We removed 17 DGs at $z \lesssim 2$ among the 1,000 MC-best models for BBG\_9 to calculate the confidence ranges. }
\tablenotetext{a}{All MC-best solutions fell in the same value.}
\end{table*}

%\begin{figure}[httb]
\begin{figure*}[]
\begin{center}
\includegraphics[width=1.0\linewidth, angle=0]{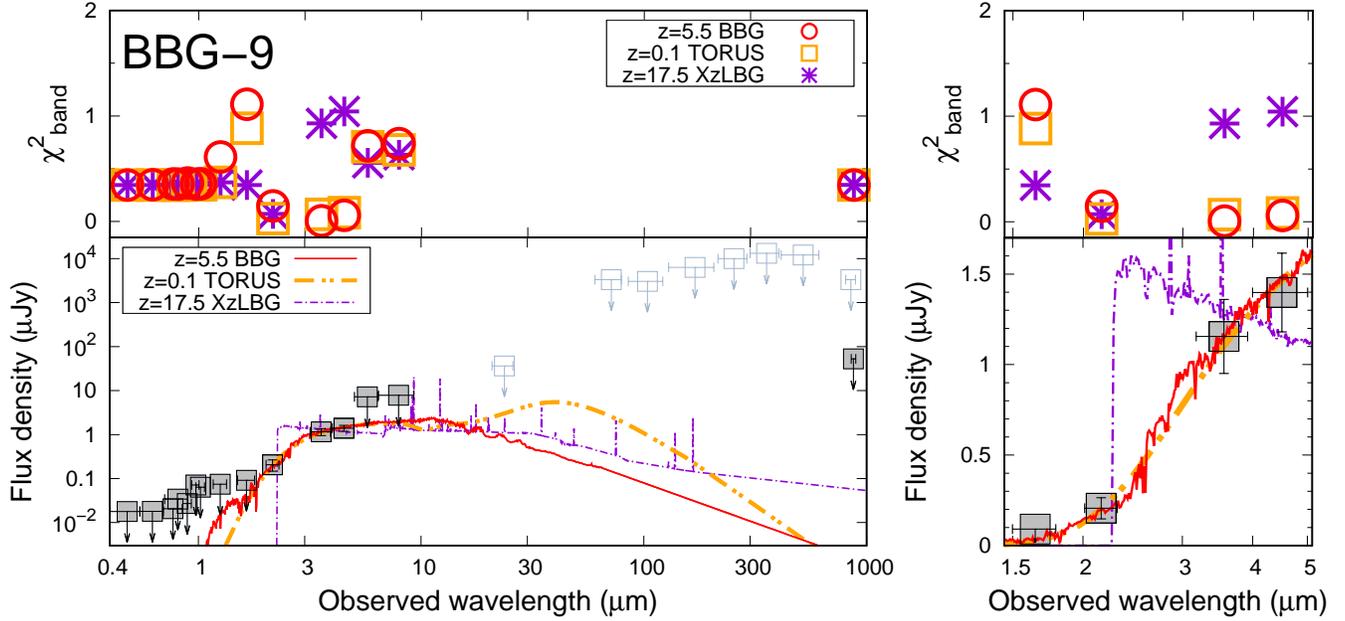}
\end{center}
\caption{The observed SED of BBG\_9 is shown in the bottom left panel, where filled (open) squares correspond to the observed photometry used (excluded) in the template fitting. For the non-detection bands, the $2\,\sigma$ limiting fluxes are set as upper limits, indicated by arrows. The best-fit spectra from the fittings with Galaxy, AGN, and XzLBG group templates are superposed. In the top left panel, $\chi^2$ values of the individual bands for each template are shown. The right panels are the same as the left panels but a zoom-up in the NIR wavelength range.  \label{fig:ALMABBG9_SED}}
\end{figure*}

%\begin{figure}[httb]
\begin{figure*}[]
\begin{center}
\includegraphics[width=1.0\linewidth, angle=0]{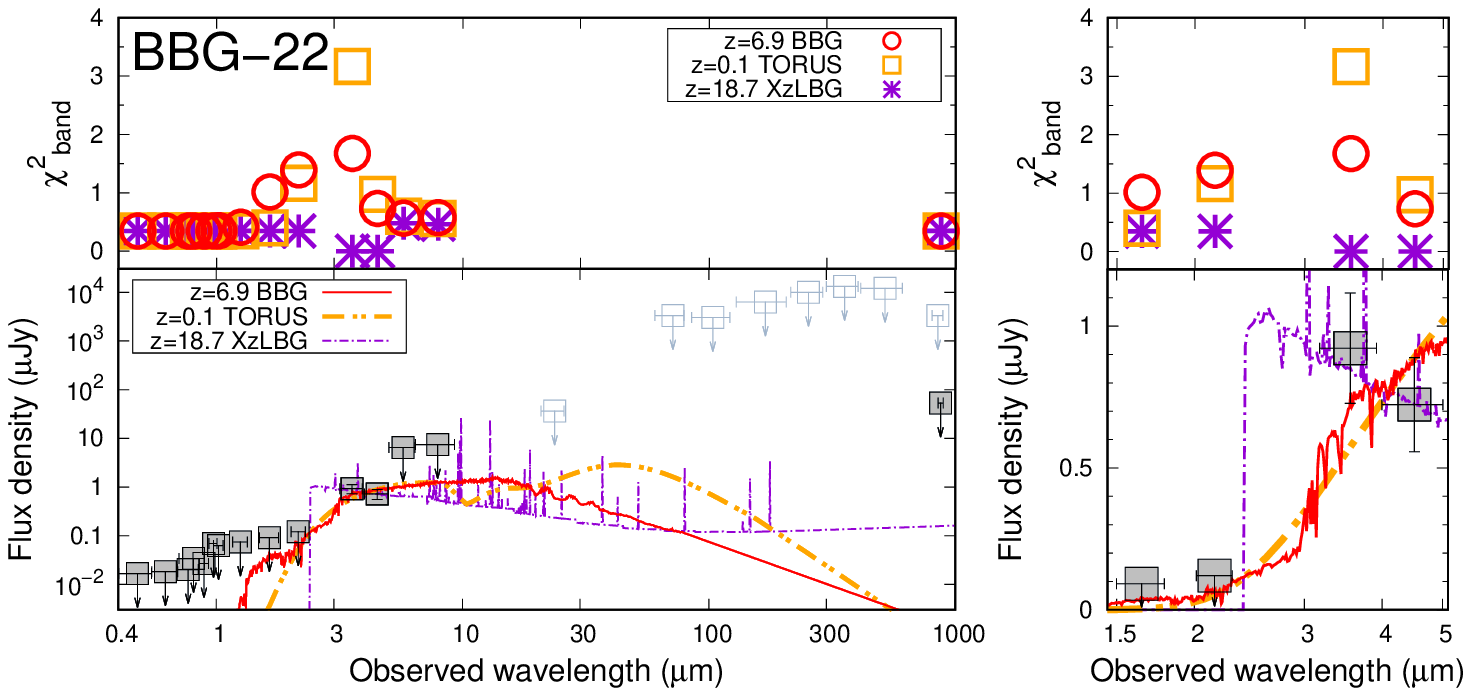}
\end{center}
\caption{Same as Figure~\ref{fig:ALMABBG9_SED} but for BBG\_22. \label{fig:ALMABBG22_SED}}
\end{figure*}

%\begin{figure}[httb]
\begin{figure*}[]
\begin{center}
\includegraphics[width=1.0\linewidth, angle=0]{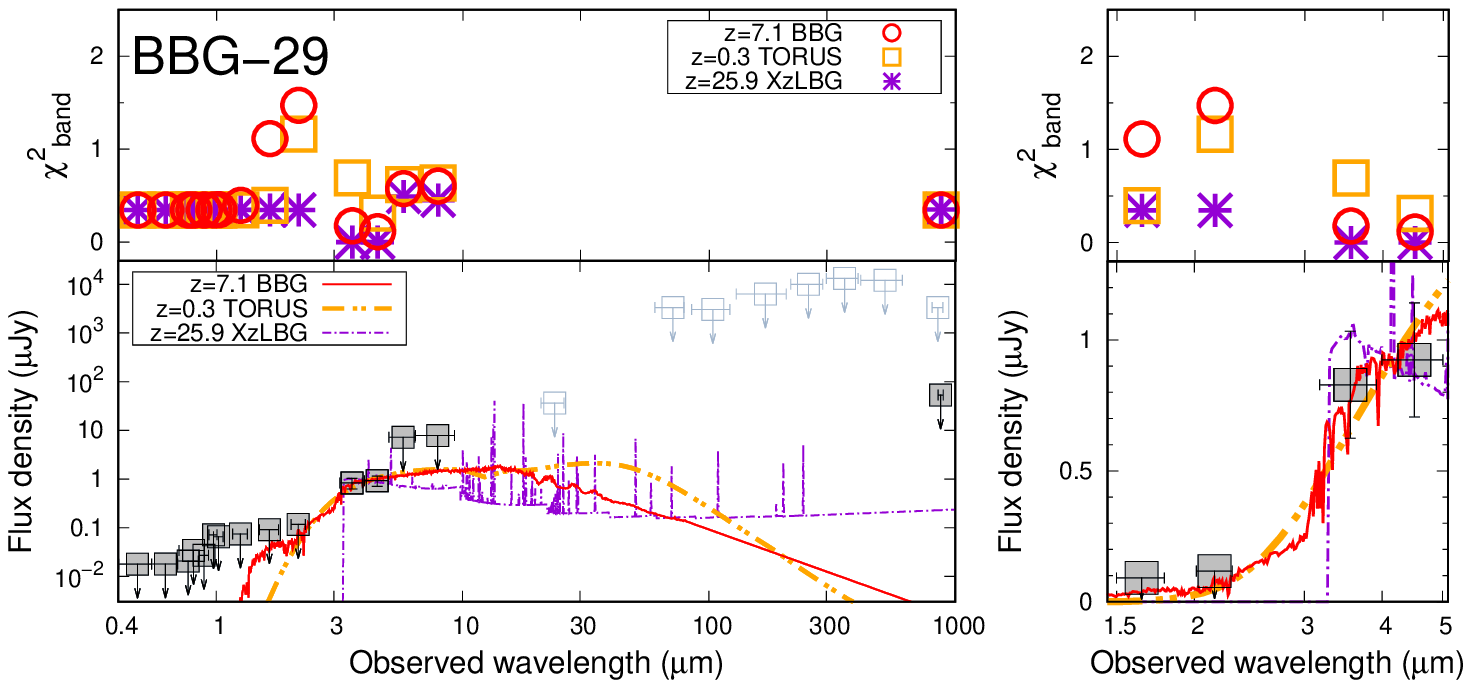}
\end{center}
\caption{Same as Figure~\ref{fig:ALMABBG9_SED} but for BBG\_29. \label{fig:ALMABBG29_SED}}
\end{figure*}

\subsubsection{Fitting with either Galaxy or AGN or XzLBG template group}\label{sec:Fit4ALMABBG_fid}

% With Galaxy group templates
First, we perform the SED fitting with the Galaxy group templates (Table~\ref{tb:SEDfitset}). As a result of 1,000 MC runs, we found that the BBG models at $5 \lesssim z \lesssim 8$ are significantly favored for all of the three BBG candidates. Figure~\ref{fig:MChist} shows the probability distributions of the five fitting parameters and the stellar mass of the 1,000 MC-best templates. The MC-best models are massive ($M_*\sim5\times10^{10}$ M$_\odot$), dust-poor ($A_V<0.2$), metal-enriched ($\sim Z_\odot$), and old (0.7--1\,Gyr) galaxies at $5\lesssim z \lesssim 8$. Their SFH is extremely bursty ($\tau_{\rm SFR} = 0.03$\,Gyr), which leads to almost zero SFRs at the observed epoch\footnote{Small SFR is also expected from the observed photometry in the FIR and optical bands. We estimated the SFR upper limit from the ALMA Band\,7 flux upper limit assuming a modified black body with a dust temperature of $T_d = 35$\,K and a conversion factor from $L_{\rm IR}$ to SFR \citep{Madau+14}. All three BBG candidates have SFR $\lesssim 10\,M_{\odot}$\,yr$^{-1}$ ($3 \sigma$). Almost the same constraint is obtained from the observed flux upper limit in the $Y$-band that roughly corresponds to the rest-frame UV wavelength at the best-fit redshifts. The SFRs of the BBG candidates are an order smaller than those of $z \sim 6$ star-forming galaxies on the main-sequence \citep{Speagle+14}) with similar stellar masses ($\sim 5 \times 10^{10}$\,M$_\odot$).}. The above SED properties are similar to those of local passive galaxies (e.g., \citealt{Cox00,Phillipps05}). We note that none of the 1,000 MC realizations result in the DG/DNLE solutions except for BBG\_9. In BBG\_9, our MC realizations result in DG solutions on rare occasions (1.7\,\% occurrence rate)\footnote{The DG solutions for BBG\_9 have old stellar populations with $T_{\rm age} \sim 3$\,Gyr, a short star-formation time-scale of $\tau_{\rm SFH} = 0.06$\,Gyr, and dust attenuation of $A_V>3$. In fact, these passive DG solutions are found around the boundary of the forbidden area of the SFR-$A_V$ plane defined in Appendix~\ref{sec:Av2SFR}.}. 
%which are required to reconcile the observed very faint FIR emission and the red $K - [3.6]$ color without any spectral break for the redshifts. 
%Such dusty but passive nature is very different from commonly observed dusty star-forming galaxies at $z \sim 2$.
We summarize the physical properties of the best-fit BBG models in Table~\ref{tb:fitprop}, where the uncertainties are derived from the MC realizations excluding the DG solutions for BBG\_9.

% with AGN group templates
Next, we performed SED fitting with the AGN group templates (Table~\ref{tb:SEDfitset}). We found that the $z \sim 0$ type-2 TORUS models are selected as the best-fit templates for all the three candidates. From 300 MC iterations of the SED fitting, 68\,\% confidence ranges on their redshifts are $z \lesssim 0.3$ for all three objects. Unfortunately, the type-2 TORUS models cannot be completely ruled out even with our deep ALMA Band\,7 constraints. This is because the type-2 TORUS dust emission is peaked at $\lambda \sim 40\,\mu$m and becomes very faint in the longer FIR regime (Figure~\ref{fig:modelspec}). In the MC-best TORUS models, the bolometric luminosity emitted by the central AGN, which is one of the model parameters \citep{Fritz+06}, is $L_{\rm bol} = (2.7^{+1.1}_{-0.7}) \times$, $(3.0^{+14.1}_{-1.4}) \times$, and $(3.3^{+3.2}_{-1.1}) \times 10^{41}$\,erg\,s$^{-1}$ for BBG\_9, 22, and 29, respectively. 

% with XzLBG group templates
Finally, SED fitting with the XzLBG group templates results in the MC-best templates at $17 < z < 20$, $19 < z < 26$, and $19 < z < 27$ (68\,\% confidence interval from 300 MC runs) for BBG\_9, 22, and 29, respectively. All the 300 MC realizations result in $A_V = 0$. Their stellar masses and SFR are as large as $3 \times 10^{9}\,M_{\odot} \lesssim M_{*} \lesssim 3 \times 10^{10}\,M_{\odot}$ and $200\,M_{\odot}\,{\rm yr}^{-1} \lesssim {\rm SFR} \lesssim 2000\,M_{\odot}\,{\rm yr}^{-1}$ even at extremely high-$z$. 

% Comparison between three
Figures~\ref{fig:ALMABBG9_SED}, \ref{fig:ALMABBG22_SED}, and \ref{fig:ALMABBG29_SED} show the best-fit spectra from the three template groups for the three BBG candidates. All the three types of templates apparently agree well with the observed SEDs with the similar fitting $\chi^2$ values both in total and in each band. We confirmed that the above results do not significantly change if we change the dust attenuation law to the SMC or MW law. 

% Incl. new HSC S18A
As mentioned in \S\ref{sec:StrictSelection}, BBG\_22 and BBG\_29 are possibly detected with $> 2\sigma$ in the HSC $i$- or $z$-bands. We also performed SED fitting for the brightest photometric measurements in the HSC bands (\S\ref{sec:StrictSelection}) with the same parameter setting as above. The best-fit $\chi^2$ values are as large as 20--30. This is because no template can consistently reproduce all the $K - [3.6]$, $[3.6] - [4.5]$ colors, non-detection in the ALMA Band\,7, and detection in the $i$/$z$-bands. We further attempted to fit the SEDs including the possible $i$/$z$ detections with two stellar populations (c.f., \citealt{Hashimoto+18a,Tamura+19}). The best-fit solution was provided by a combination of an old passive model with $T_{\rm age} \approx 1$\,Gyr and a young star-forming model with ${\rm SFR} \lesssim 10\,M_{\odot}$\,yr$^{-1}$ at $5 \lesssim z \lesssim 5.5$. Because the best-fit stellar masses of the old passive components are very similar to those of the best-fit BBG models obtained above (Table~\ref{tb:fitprop}), the following discussion about stellar mass density (SMD) and SFRD is not sensitive to whether the observed three galaxies are marginally detected or not in the HSC $i$- or $z$-bands.

\subsubsection{Fitting with composite templates of Star+Nebular+Dust and TORUS models}\label{sec:Fit4ALMABBG_sup}

%\begin{figure}[httb]
\begin{figure}[]
\begin{center}
\includegraphics[width=1.0\linewidth, angle=0]{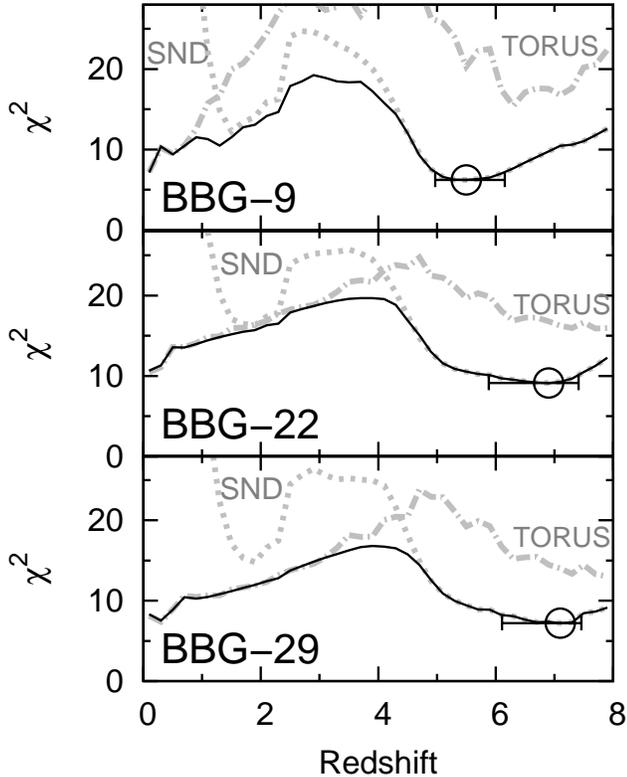}
\end{center}
\caption{SED fitting $\chi^2$ as a function of redshift for the three BBG candidates without ALMA detection. Dotted, dot-dashed, and solid lines correspond to the $\chi^2$ distributions in the fitting with the Star+Nebular+Dust, TORUS, and their composite templates (see text), respectively. The circles with error-bars show the least $\chi^2$ point and the range of $\Delta \chi^2 \leq 1$ in the fitting with the Star+Nebular+Dust and TORUS composite templates.  \label{fig:z2chi2_Star+Nebular+Dust+TORUS}}
\end{figure}

In \S\ref{sec:Fit4ALMABBG_fid}, we treated the galaxy and AGN templates separately because the connection between a galaxy and an AGN is not trivial and the number of the combinations is too large. In contrast, as every AGN is part of a galaxy, it is worthwhile to fit the observed SEDs with combined templates of Star+Nebular+Dust models and AGN TORUS models \citep{Fritz+06,Feltre+12}. We generated 5,883,840 combined templates of the Star+Nebular+Dust models and TORUS models at $0 < z < 8$. Although the Star+Nebular+Dust models we used here have the same parameter coverage as in the Galaxy group (Table~\ref{tb:SEDfitset}), we reduced the parameter steps for $\tau_{\rm SFR}$, $T_{\rm age}$, and $A_V$. The TORUS models at each redshift are reduced from 24,000 to 10 representatives that are the same as those used in the code ``SED3FIT''\footnote{http://steatreb.altervista.org/alterpages/sed3fit.html} \citep{Berta+13a}. The TORUS spectra are scaled to the Star+Nebular+Dust models with a free parameter $L_{\rm bol}^{\rm TORUS} / M_{*}^{\rm Star+Nebular+Dust}$. Taking account of a relation between BH mass and stellar mass of the host galaxy ($10^{-5} \lesssim M_{\rm BH} / M_{*} \lesssim 10^{-2}$; \citealt{Reines+15}) and observations of the Eddington ratio ($10^{-3}$\,erg\,s$^{-1}$\,$M_{\odot}^{-1}$ $< L_{\rm bol} / (1.25 \times 10^{38} \times M_{\rm BH}) < 10$\,erg\,s$^{-1}$\,$M_{\odot}^{-1}$; \citealt{Woo+02}), we set the parameter range as wide as $1.25 \times 10^{30}$\,erg\,s$^{-1}$\,$M_{\odot}^{-1}$ $\le L_{\rm bol}^{\rm TORUS} / M_{*}^{\rm Star+Nebular+Dust} \le 1.25 \times 10^{37}$\,erg\,s$^{-1}$\,$M_{\odot}^{-1}$ with a step of $1.0$ in the common logarithmic scale. 

The resulting $\chi^2$ values as a function of redshift are shown in Figure~\ref{fig:z2chi2_Star+Nebular+Dust+TORUS}, where those from the original Star+Nebular+Dust or TORUS only templates are also superposed. We found that the local $\chi^2$ minima at high and low redshifts are achieved by templates dominated by the Star+Nebular+Dust or TORUS models, respectively. Templates equally contributed by the Star+Nebular+Dust and TORUS models do not show better fit than either of them for the observed SEDs. Therefore, we regarded the results in \S\ref{sec:Fit4ALMABBG_fid} as the fiducial ones in the following discussion.

\section{Discussion}\label{sec:discussion}

In this section, we focus on the three BBG candidates that are not detected in ALMA Band\,7. 
First, we discuss the possible contamination to the BBG candidates (\S\ref{sec:disc_contami}). Then, assuming all the three candidates are real BBGs at $z \ga 5$, we estimated their cosmic SMD and discussed the SFRD of their progenitors (\S\ref{sec:progenitor}).

\subsection{Possibility of contamination}\label{sec:disc_contami}

\subsubsection{AGNs at $z \sim 0$} \label{sec:disc_z0AGN}

Our SED fitting analyses reveal that heavily obscured AGN torus templates (Type-2 TORUS) at $z \sim 0$ give as good fits as the BBG templates. In contrast, their SEDs are very unusual. Compared to the observed dusty AGN population \citep{Polletta+07,Rigopoulou+09}, the type-2 TORUS models show an order of magnitude larger flux density ratio of the rest-frame NIR ($\lambda \sim 2\,\mu$m) to FIR ($\lambda \sim 300\,\mu$m). 
%Such exotic "bared type-2 TORUS" AGNs have never been observed so far in the real Universe.
At the Eddington accretion rate, the AGN bolometric luminosity is given by the Eddington luminosity, $L_{\rm Edd} = 1.25 \times 10^{38} \times M_{\rm BH} / M_\odot$\,erg\,s$^{-1}$, where $M_{\rm BH}$ is the BH mass. Observationally, it is known that the AGN bolometric luminosity ranges over $0.001 \times L_{\rm Edd} < L_{\rm bol} < 10 \times L_{\rm Edd}$ \citep{Woo+02}. Assuming this wide range of the Eddington ratio, we estimated the BH mass to be $240 \lesssim  M_{\rm BH} / M_\odot \lesssim 2.4 \times 10^6$ from a typical bolometric luminosity of the best-fit TORUS models of $L_{\rm bol} \sim 3 \times 10^{41}$\,erg\,s$^{-1}$ (\S\ref{sec:Fit4ALMABBG_fid}). If the TORUS solution is the case for our BBG candidates, they may be the lowest mass AGN BHs observed so far \citep{Baldassare+15,Bentz+15}.  
%Such low mass AGNs are unique and interesting even in the local Universe.

Furthermore, in the TORUS solution, emission from the host galaxy should be very faint at all wavelengths by definition (see \S\ref{sec:tempAGN}). To constrain properties of the host galaxies, we revisited SED fitting with the combined templates of the Star+Nebular+Dust and TORUS models (\S\ref{sec:Fit4ALMABBG_sup}). We examined the combined templates whose $\chi^2$ and redshift are similar to those of the best-fit $z \sim 0$ TORUS solutions. Their SEDs were almost dominated by the TORUS model templates with negligible contribution by the Star+Nebular+Dust templates, yielding a stellar mass of the host galaxies as $M_*^{\rm host} \lesssim 2 \times 10^6\,M_\odot$. 

The estimated $M_{\rm BH}$ and $M_*^{\rm host}$ give the relatively high BH-to-total stellar mass ratio ranging from $10^{-4}$ to 1, of which only the lower boundary is consistent with an observed scaling relation \citep{Reines+15}. Such an AGN with a very low-mass BH hosted by a relatively low mass galaxy that contributes little to the whole SED (``low-mass naked'' AGN) seems unlikely, while we reserve the complete rejection of the AGN solutions in the future.

\subsubsection{LBGs at $z \gtrsim 17$}

The XzLBGs at $z \gtrsim 17$ cannot be ruled out from the SED analyses. Especially for BBG\_22, the observed blue $[3.6] - [4.5]$ color prefers the XzLBG model rather than the BBG and TORUS models (Figure~\ref{fig:ALMABBG22_SED}). The observed three BBG candidates had $[3.6] \sim 1\,\mu$Jy that corresponds to absolute rest-UV magnitudes of $M_{\rm UV} \sim -24.5$\,mag at $z \sim 20$. Even with a very optimistic assumption that the UV luminosity function (UVLF) does not evolve beyond $z = 10$ \citep{Bouwens+15}, an expected number of the XzLBGs as bright as the observed objects is $\ll 1$ in the survey volume corresponding to $17 \leq z \leq 27$. Therefore, this possibility is also unlikely, although this case is highly interesting.

\subsection{Stellar mass density of the $z \sim 6$ BBGs}\label{sec:SMD}

Compared to the DGs, AGNs, and XzLBGs mentioned in the previous sections, the BBG solutions at $z \sim 6$ may be physically acceptable for the three objects. The metallicity of the best-fit BBG model in the SED fitting is already the solar value in the $z \sim 6$ Universe. This is consistent with the chemical evolution model \citep{Asano+13a} that predicts that matured galaxies with $T_{\rm age} \gtrsim 0.3$\,Gyr and $M_{*} \gtrsim 10^{10}\,M_{\odot}$ can enrich their metallicity as high as the solar level \citep{Tamura+19}. 

The passive nature of the $z \sim 6$ BBG model seems difficult to be explained within the current theoretical framework of galaxy formation. This is because frequent galaxy interactions, gas supply into galaxies from the large-scale structure, and stellar feedback induce stochastic star-formation \citep{Trebitsch+17,Hopkins+18,Ma+18,Ceverino+18}. We search for galaxies with similar stellar mass satisfying our BBG color criteria in a $\sim 1$\,Gpc$^3$ box of a semi-analytic model \citep{Makiya+16}, resulting in no such counterpart at $z > 5$. In contrast, the prominent Balmer break is observationally confirmed in the $z = 9.1$ galaxy \citep{Hashimoto+18a}, suggesting a passive phase lasting for $\sim 100$\,Myr or longer in galaxies even in the very early Universe. 

Assuming all of the three BBG candidates without ALMA detections to be passive galaxies at $4.8 \leq z \leq 7.8$ and correcting for the detection completeness ($0.53$ at $[3.6] \sim 24$\,mag; Figure~\ref{fig:completeness}), we estimate the number density of the BBGs as $n_{\rm BBG} = (4.9_{-2.7}^{+4.8}) \times 10^{-7}$\,Mpc$^{-3}$ (comoving). Adopting the best-fit stellar mass and the uncertainties derived from the SED analysis, we obtained the SMD of $(2.4_{-1.3}^{+2.3}) \times 10^4\,M_\odot$\,Mpc$^{-3}$. 

Figure~\ref{fig:SFRD} (top panel) shows our SMD estimate as a function of redshift in conjunction with those from literature for both passive galaxies \citep{Muzzin+13c,Straatman+14,Davidzon+17} and star-forming galaxies \citep{Muzzin+13c,Duncan+14,Grazian+15,Song+16,Davidzon+17,Bhatawdekar+19,Kikuchihara+19}. The SMDs in the previous works except for those in \citet{Straatman+14} are estimated by integrating the stellar mass functions (SMFs) down to $M_{*} = 10^{8}\,M_{\odot}$. In this work and \citet{Straatman+14}, for passive galaxies at $z \sim 6$ and $\sim 4$, respectively, however, the limited sample sizes prevented the authors from constructing SMFs. The estimated SMDs in \citet{Straatman+14} and this work are contributed only by massive galaxies down to the observational mass limits of $\sim 2 \times 10^{10}$ and $\sim 4 \times 10^{10}\,M_{\odot}$, respectively. This difference in the mass limits does not affect the SMDs for passive galaxies so much. This is because the SMFs for passive galaxies at $z \lesssim 4$ show rapid decrease at $M_{*} \lesssim 3 \times 10^{10}\,M_{\odot}$ \citep{Muzzin+13c,Davidzon+17} and then the less massive galaxies do not contribute to the SMDs. Our SMD at $z \sim 6$ is broadly consistent with the decreasing trend of passive galaxies from $z = 0$ to $4$. The fraction contributed by the BBGs in the total SMD including star-forming galaxies at $z \sim 6$ is only $\sim 1$ percent.

\subsection{Cosmic star-formation activity at $z \ga 14$}\label{sec:progenitor}

\begin{figure*}[]
\begin{center}
\includegraphics[width=1.0\linewidth, angle=0]{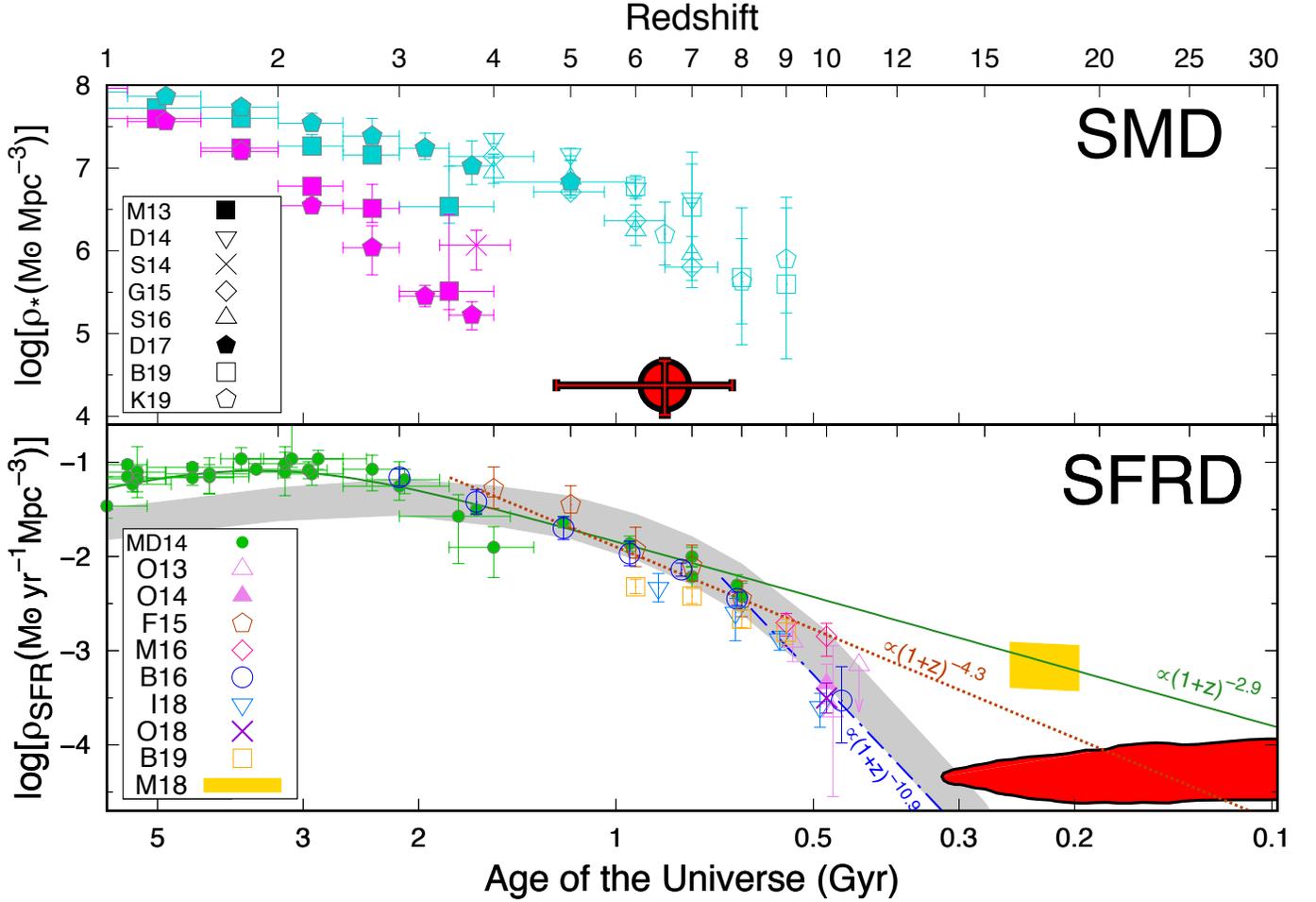}
\end{center}
\caption{Evolution of the stellar mass density (SMD: top) and the star-formation rate density (SFRD: bottom) along the cosmic history (see top axis for the corresponding redshift). For these plots, we assumed all three BBG candidates without ALMA detection to be real passive galaxies at $z \sim 6$. In the top panel, the SMD of our BBG sample at $z \sim 6$ (red circle) is shown in conjunction with those of star-forming (cyan symbols) and passive (magenta symbols) galaxies at lower redshifts from the literature (M13: \citealt{Muzzin+13c}, D14: \citealt{Duncan+14}, S14: \citealt{Straatman+14}, G15: \citealt{Grazian+15}, S16: \citealt{Song+16}, D17: \citealt{Davidzon+17}, B19: \citealt{Bhatawdekar+19}, and K19: \citealt{Kikuchihara+19}). The vertical error-bar associated with our BBG data corresponds to a $1\,\sigma$ uncertainty propagated from the Poisson error \citep{Gehrels86} for the BBG number and the SED fitting uncertainty for the stellar mass. The horizontal error bar shows the redshift range expected from our BBG color selection. In the bottom panel, the red shade corresponds to the SFRD expected from the progenitors of the $z \sim 6$ BBGs at a $99.7$\,\% confidence level ($3\,\sigma$). The SFRD measurements at $z \lesssim 10$ are collected from the literature (MD14: \citealt{Madau+14}, O13: \citealt{Oesch+13b}, O14: \citealt{Oesch+14}, F15: \citealt{Finkelstein+15a}, M16: \citealt{McLeod+16}, B16: \citealt{Bouwens+16c}, I18: \citealt{Ishigaki+18}, O18: \citealt{Oesch+18a}, and B19: \citealt{Bhatawdekar+19}). All of them at $4 \lesssim z \lesssim 10$ are estimated by integrating the UV LFs down to $M_{\rm UV} = -17$\,mag. The SFRD estimated at $z \sim 17$ from an observed global 21\,cm absorption trough (M18: \citealt{Madau18}, \citealt{Bowman+18}) is also shown in yellow. The functional fit to the MD14 data, which is proportional to $(1 + z)^{-2.9}$ at high-$z$ \citep{Madau+14}, is superposed by the solid line. Two other power law functions supporting an accelerated evolution at $z \ga 8$ ($\rho_{\rm SFR} \propto (1+z)^{-10.9}$; \citealt{Oesch+14}) and a smooth evolution from lower redshift ($\rho_{\rm SFR} \propto (1+z)^{-4.3}$; \citealt{Finkelstein+15a}) are shown by dot-dashed and dotted lines, respectively. The SFRD derived assuming a universal relation among the halo mass, SFR, and dark matter accretion rate \citep{Harikane+18a} is also superposed by the grey shade in its $1\,\sigma$ uncertainty. All the SMD and SFRD measurements from the literature are corrected for the stellar IMF and the cosmological model to match those in this work. \label{fig:SFRD}}
\end{figure*}

The small but non-zero number of BBGs suggests the star-formation activity by their progenitors at $z \ga 10$, several hundred million years or more before the observed epoch of $z \sim 6$. In the following section, we discuss the cosmic star formation activity by such progenitors, assuming that all the three candidates are real BBGs. 

The number density of the star-forming progenitors can be different from that of the BBGs because the progenitors are observable only during their star-forming phase ($T_{\rm SF}$) while the descendant BBGs are always observable once the Balmer break matures. Assuming the BBG observable time duration as $T_{\rm BBG} = T_{\rm age} - T_{\rm SF} - 0.2\,{\rm Gyr}$ with $T_{\rm SF} = 2 \times \tau_{\rm SFH}$, where the 0.2\,Gyr is required to develop the Balmer break, we obtained the number density of the star-forming progenitors as $n_{\rm prog} = n_{\rm BBG} \times T_{\rm SF} / T_{\rm BBG} \approx 5.6 \times 10^{-8} (T_{\rm SF} / 60\,{\rm Myr} )$\,Mpc$^{-3}$ (comoving). We should note here that $T_{\rm SF}$ is not strongly constrained by the SED fitting. The observed Balmer break demands a sufficiently long $T_{\rm age}$. As a result, $T_{\rm SF}$ in the exponential SFH has to be short because of the limited cosmic time at $z \sim 6$. It is possible that an SFH having a certain star-formation before the exponentially declining one provides a good fit to the BBGs' SED and allows a bit longer $T_{\rm SF}$ (but still limited by the short cosmic age). In this case, $n_{\rm prog}$ becomes larger than the above value accordingly. 
 
%In estimation of SFRD owed by the star-forming progenitors, we need to analyze the progenitors' SFR also taking account of the observable time difference between the BBGs and their progenitors. 
We estimate the SFRD as follows:
\begin{eqnarray}
\rho_{\rm SFR} = <{\rm SFR}> \times n_{\rm prog} = \sum_{i = 1}^{3} {\rm SFR}_{i}^{\rm prog} \times \frac{1}{V_{\rm eff}} \times \frac{T_{{\rm SF},i}}{T_{{\rm BBG},i}} \label{eq:progSFRD1}\,,
\end{eqnarray}
where $<{\rm SFR}>$ is the average for the three objects, index $i$ corresponds to each BBG, and $V_{\rm eff}$ is the effective survey volume at $4.8 < z < 7.8$ corrected for the detection incompleteness. Each progenitor's SFR (SFR$_{i}^{\rm prog}$) and star-forming duration ($T_{{\rm SF},i}$) are very sensitive to the SFH functional shape in the SED fitting. However, it can be approximated as 
\begin{eqnarray}
\rho_{\rm SFR} \approx \sum_{i = 1}^{3} \frac{M_{*,i}^{\rm BBG}}{T_{{\rm SF},i}} \times \frac{1}{V_{\rm eff}} \times \frac{T_{{\rm SF},i}}{T_{{\rm BBG},i}} \approx \frac{\rho_{*}^{\rm BBG}}{<T_{\rm BBG}>}\label{eq:progSFRD2}\,,
\end{eqnarray}
where $<T_{\rm BBG}>$ is an average for the three objects and $\rho_{*}^{\rm BBG}$ is the SMD of the $z \sim 6$ BBGs estimated in the previous section. This suggests that we can obtain the SFRD independent of the uncertain SFH and $T_{\rm SF}$. Therefore, the obtained SFRD may be as robust as the SMD because $M_{*}$ and $T_{\rm BBG}$ are relatively well constrained by the observed IRAC flux density and $[3.6] - [4.5]$ color (Balmer break strength).

Practically, we estimated the progenitors' SFRD adopting Equation~(\ref{eq:progSFRD1}) for the BBG models obtained from the SED analysis. Here, we need to specify the timing to pick the SFR value and its redshift. We chose the model age same as the star-formation timescale ($T_{\rm age} = \tau_{\rm SFH}$), and this choice does not affect the obtained SFRD as explained above. We made $10^6$ combinations of three MC-best BBG models from the randomly selected 100 MC realizations for each object, from which we evaluated the confidence range on the SFRD. The progenitors' redshifts were simply averaged for each combination. The resultant $99.7$\,\% confidence range ($3\,\sigma$) on the progenitors' SFRD and redshift is shown in Figure~\ref{fig:SFRD} (bottom panel). Our estimate, $2.4 \times 10^{-5} \lesssim {\rm SFRD} / M_\odot\,{\rm yr}^{-1}\,{\rm Mpc}^{-3} \lesssim 1.2 \times 10^{-4}$ at $z \ga 14$, should be regarded as a lower limit because not all star-forming galaxies at $z \ga 14$ evolve into passive galaxies at $z \sim 6$. We also note that a similar estimate is obtained from Equation~(\ref{eq:progSFRD2}): $\rho_{\rm SFR} \approx (2.4 \times 10^4\,M_{\odot}\,{\rm Mpc}^{-3}) / (0.56\,{\rm Gyr}) = 4.3 \times 10^{-5}\,M_{\odot}\,{\rm yr}^{-1}\,{\rm Mpc}^{-3}$, where $<T_{\rm BBG}> = 0.56$\,Gyr. 

For comparison, we collected SFRD measurements at $z \lesssim 10$ from the literature \citep{Oesch+13b,Madau+14,Oesch+14,Finkelstein+15a,McLeod+16,Bouwens+16c,Ishigaki+18,Oesch+18a,Bhatawdekar+19}. Among these, the measurements at $4 \lesssim z \lesssim 10$ were converted from UV luminosity densities derived by integrating the UV LFs down to a certain luminosity or magnitude limit ($L_{\rm lim}$ or $M_{\rm lim}$). To correct for the different $M_{\rm lim}$ and conversion factors from UV luminosity densities to SFRD adopted in the previous works, we re-integrated their UV LFs down to $M_{\rm lim} = -17$ and multiplied the conversion factor of \citet{Madau+14}. The measurements of \citet{Madau+14} were actually not re-estimated because their adopted $L_{\rm lim} = 0.03 L^{*}$, where $L^{*}$ is the characteristic luminosity of the UV LF, corresponds to $M_{\rm lim} = -17$\,mag at $z \sim 3$ \citep{Reddy+09}. All of the literature measurements are also corrected for the IMF and the cosmological parameters to match with those adopted in this work. 

The SFRDs from the previous works after the above corrections are shown in Figure~\ref{fig:SFRD}. In addition to the direct measurements at $z \lesssim 10$, we also put an SFRD estimate at $z \sim 17$ (yellow shade in Figure~\ref{fig:SFRD}) based on the UV luminosity density of \citet{Madau18} using an SFR conversion factor \citep{Madau+14} corrected for the IMF. The UV luminosity density of \citet{Madau18} was estimated to reproduce a tentative detection by EDGES collaboration of global 21\,cm absorption trough imprinted in the cosmic microwave background (CMB) spectrum \citep{Bowman+18} under the assumption that the 21\,cm signal is activated by extremely metal-poor stellar systems. 

We now revisit the frequently debated topic on SFRD evolution at $z \ga 8$. In Figure~\ref{fig:SFRD}, four possible evolutionary trends are also shown. Two support a smooth evolution from $z \sim 5$ to $10$: the functional-fit to the measurements of \citet{Madau+14} that is proportional to $(1 + z)^{-2.9}$ at $z > 3$, and a bit steeper power-law function with the slope $\alpha = -4.3$ proposed by \citet{Finkelstein+15a}. The others support a rapid SFRD decline at $z \ga 8$: a power-law function with the slope $\alpha = -10.9$ \citep{Oesch+14}, and an expected evolution assuming no redshift dependence on the relation among the halo mass, SFR, and dark matter accretion rate \citep{Harikane+18a}. Our SFRD estimate at $z \ga 14$ seems consistent with the smooth SFRD evolution.
If our SFRD is a lower limit as discussed above, it is also consistent with the $z \sim 17$ estimate based on the EDGES detection of the 21\,cm absorption signal \citep{Madau18,Bowman+18} that is on the extrapolation of the \citet{Madau+14} best-fit function. In contrast, the rapid decrease in the SFRD may not be consistent with our estimate. Because the rapid decline can be interpreted by the number density evolution of dark matter halos \citep{Oesch+18a,Harikane+18a}, our relatively high SFRD may indicate a higher star-formation efficiency in the halos at $z \ga 10$. 

Individual progenitors of the BBGs should have produced their stellar mass of $\sim 5 \times 10^{10}\,M_\odot$ by $z \sim 14$. The number density of the star-forming progenitors, $n_{\rm prog} \approx 5.6 \times 10^{-8}$\,Mpc$^{-3}$, implies their halo mass to be $\sim 10^{11}\,M_\odot$ ($\sim 10^{12}$) at $z \sim 17$ ($z \sim 11$) \citep{Mo+02}. Their stellar-to-halo mass ratios (SHMRs) are expected to be $\sim 0.5$ ($\sim 0.05$) at $z \sim 17$ ($z \sim 11$). These are much larger than SHMRs measured at lower redshift (e.g., \citealt{Leauthaud+12,Behroozi+13d}). Our measurement is hard to be explained if we linearly extrapolate a redshift evolution of SHMRs \citep{Finkelstein+15b,Harikane+16} up to $z \sim 20$. The very high SHMRs again suggest an unexpectedly high star-formation efficiency and/or low feedback efficiency in the BBG progenitor halos at the pre-reionization epoch. This problem should be resolved by theoretical works in the future.

\section{Conclusion}

In this study, we searched for passively evolving galaxies whose SED is characterized by the prominent Balmer break. 
The effective survey area is $0.41$\,deg$^2$ in the COSMOS field. Using the rich imaging data set available in the entire survey field, we photometrically identified six candidate BBGs. We performed follow-up observations with ALMA Band\,7 for these BBG candidates. Three among the six candidate BBGs were detected in dust continuum emission. The remaining three candidates not detected with ALMA are promising BBG candidates. 
Through comprehensive SED analyses with a large template set of galaxy and AGN models, we obtained the following results and implications. 
\begin{itemize}
%\item For the BBG candidates with ALMA detections, their SEDs are well explained by the dusty galaxy models with $L_{\rm IR} \sim 10^{12}\,L_{\odot}$ at $z \ga 4$. While their redshifts are not well constrained, we conclude that they are high-$z$ ULIRGs. 
\item The three BBG candidates not detected with ALMA can be considered as the most likely BBGs at $z \sim 6$. The best-fit galaxy models for their SEDs have the following properties: $5 < z < 8$; $M_{*} \sim 5 \times 10^{10}\,M_{\odot}$; in the inactive star-formation phase for $\ga 0.7$\,Gyr; no dust attenuation/emission; metal-enriched as a solar level. 
\item The cosmic SMD estimated from the three most likely BBG candidates at $5 \lesssim z \lesssim 8$ is $\rho_* = (2.4^{+2.3}_{-1.3}) \times 10^4\,M_{\odot}$\,Mpc$^{-3}$. This is consistent with the decreasing trend observed at $z < 4$. 
\item The onset of star-formation in the most likely BBG candidates should be at $z \ga 14$. The cosmic SFRD contributed by the progenitors is expected to be $2.4 \times 10^{-5} \lesssim \rho_{\rm SFR} / M_{\odot}\,{\rm yr}^{-1}\,{\rm Mpc}^{-3} \lesssim 1.2 \times 10^{-4}$ with $99.7$\,\% confidence ($3\,\sigma$). This SFRD estimate is less sensitive to SFHs assumed in the SED fitting analyses. The SFRD contributed by the progenitors of the BBGs is a lower limit of the total SFRD owed by all populations of galaxies at $z \ga 14$. Our estimate supports a smooth evolution of SFRDs from $z \sim 5$ to beyond $z \sim 10$ rather than an accelerated evolution at $z \ga 8$. 
\item In the most likely BBG sample, however, there is still possible contamination from type-2 AGNs at $z \sim 0$ with very low mass ($240\,M_{\odot} \lesssim M_{\rm BH} \lesssim 2.4 \times 10^6\,M_{\odot}$) BHs hosted by relatively low mass galaxies ($M_{*}^{\rm host} \lesssim 2 \times 10^{6}\,M_{\odot}$) that contribute little to the whole SEDs. Such low mass naked type-2 AGNs seem to be unlikely but very interesting, for which follow-up observations with future deeper NIR or MIR instruments are needed. 
\end{itemize}

While the above results are based on the best effort using the highest quality imaging data currently available, direct evidence of the BBG, i.e., spectroscopic confirmation of the Balmer break by coming {\it James Webb Space Telescope} ($JWST$; \citealt{Gardner+06}), is needed. If our BBG sample is really at $z \ga 5$, it is time to construct a new formation path for massive galaxies in the very early Universe.

%% If you wish to include an acknowledgments section in your paper,
%% separate it off from the body of the text using the \acknowledgments
%% command.
\acknowledgments

% Financial support
K. M. and A. K. I. acknowledge financial support from Japan Society for the Promotion of Science (JSPS) in KAKENHI Grant number 26287034 and 17H01114. T. H. and A. K. I. acknowledge financial support by NAOJ ALMA Scientific Research Grant Number 2016-01. Y. T. is supported by JSPS in KAKENHI Grant number 16H02166. 
% Communication
We thank M. Sawicki for discussions regarding the SED fit algorithm, M. Akiyama for discussions regarding the AGN solutions, A. Goulding for discussion in the HSC data analysis, N. Scoville for interpretation of the SED fit results, and C. H. Lee for the English corrections. We also thank I. Smail and C. C. Chen for letting us know about the detections of the previously reported BBG candidate \citep{Mawatari+16b} by SCUBA-2 and ALMA. We would like to thank Editage (www.editage.com) for English language editing. We appreciate the observatories where the data used in this work were taken. 
% Spitzer
The {\it Spitzer Space Telescope} is operated by the Jet Propulsion Laboratory, California Institute of Technology, under a contract with NASA. 
% VISTA
The UltraVISTA survey is based on data products from observations made with ESO Telescopes at the La Silla Paranal Observatory under ESO programme ID 179.A-2005 and on data products produced by TERAPIX and the Cambridge Astronomy Survey Unit on behalf of the UltraVISTA consortium. 
% ALMA
This paper makes use of the following ALMA data: ADS/JAO.ALMA \#2017.1.01259.S. ALMA is a partnership of ESO (representing its member states), NSF (USA), and NINS (Japan), together with NRC (Canada), MOST and ASIAA (Taiwan), and KASI (Republic of Korea), in cooperation with the Republic of Chile. The Joint ALMA Observatory is operated by ESO, AUI/NRAO, and NAOJ. 
% HST
The $HST$ data used in this work were obtained through the data archive at the Space Telescope Science Institute, which is operated by the Association of Universities for Research in Astronomy, Inc. under NASA contract NAS 5-26555. 
% HSC
The Hyper Suprime-Cam (HSC) collaboration includes the astronomical communities of Japan and Taiwan, and Princeton University. The HSC instrumentation and software were developed by the National Astronomical Observatory of Japan (NAOJ), the Kavli Institute for the Physics and Mathematics of the Universe (Kavli IPMU), the University of Tokyo, the High Energy Accelerator Research Organization (KEK), the Academia Sinica Institute for Astronomy and Astrophysics in Taiwan (ASIAA), and Princeton University. Funding was contributed by the FIRST program from Japanese Cabinet Office, the Ministry of Education, Culture, Sports, Science and Technology (MEXT), the Japan Society for the Promotion of Science (JSPS), Japan Science and Technology Agency (JST), the Toray Science Foundation, NAOJ, Kavli IPMU, KEK, ASIAA, and Princeton University. This paper used software developed for the Large Synoptic Survey Telescope. We thank the LSST Project for making their code available as free software at http://dm.lsst.org. The Pan-STARRS1 Surveys (PS1) have been made possible through contributions of the Institute for Astronomy, the University of Hawaii, the Pan-STARRS Project Office, the Max-Planck Society and its participating institutes, the Max Planck Institute for Astronomy, Heidelberg and the Max Planck Institute for Extraterrestrial Physics, Garching, The Johns Hopkins University, Durham University, the University of Edinburgh, Queen’s University Belfast, the Harvard-Smithsonian Center for Astrophysics, the Las Cumbres Observatory Global Telescope Network Incorporated, the National Central University of Taiwan, the Space Telescope Science Institute, the National Aeronautics and Space Administration under Grant No. NNX08AR22G issued through the Planetary Science Division of the NASA Science Mission Directorate, the National Science Foundation under Grant No. AST-1238877, the University of Maryland, and Eotvos Lorand University (ELTE), and the Los Alamos National Laboratory. The HSC-SSP data are collected at the Subaru Telescope and retrieved from the HSC data archive system, which is operated by Subaru Telescope and Astronomy Data Center at National Astronomical Observatory of Japan. 
% Herschel
$Herschel$ is an ESA space observatory with science instruments provided by European-led Principal Investigator consortia and with important participation from NASA. 
% Scuba-2
The SCUBA-2 is equipped on the James Clerk Maxwell Telescope, which is operated by the East Asian Observatory on behalf of The National Astronomical Observatory of Japan, Academia Sinica Institute of Astronomy and Astrophysics, the Korea Astronomy and Space Science Institute, the National Astronomical Observatories of China and the Chinese Academy of Sciences (Grant No. XDB09000000), with additional funding support from the Science and Technology Facilities Council of the United Kingdom and participating universities in the United Kingdom and Canada. 
% VLA
The VLA is a part of the National Radio Astronomy Observatory which is a facility of the National Science Foundation operated under cooperative agreement by Associated Universities, Inc. 
% XMM-Newton
The {\it XMM-Newton satellite} is an ESA science mission with instruments and contributions directly funded by ESA Member States and NASA. 
% Chandra
The scientific results reported in this article are based in part on observations made by the {\it Chandra X-ray Observatory}.

%% To help institutions obtain information on the effectiveness of their 
%% telescopes the AAS Journals has created a group of keywords for telescope 
%% facilities.
%
%% Following the acknowledgments section, use the following syntax and the
%% \facility{} or \facilities{} macros to list the keywords of facilities used 
%% in the research for the paper.  Each keyword is check against the master 
%% list during copy editing.  Individual instruments can be provided in 
%% parentheses, after the keyword, but they are not verified.

\vspace{5mm}
\facilities{$Spitzer$(IPAC), VISTA(ESO), Subaru(NAOJ), $HST$(STIS), $Herschel$(ESA), JCMT(EAO), VLA(NRAO), $XMM-Newton$(ESA), $Chandra$(NASA)}

%% Similar to \facility{}, there is the optional \software command to allow 
%% authors a place to specify which programs were used during the creation of 
%% the manusscript. Authors should list each code and include either a
%% citation or url to the code inside ()s when available.

\software{SExtractor \citep{BertinArnouts96}, 
             IRAF \citep{Tody86,Tody93}, \\
             PANHIT (http://www.icrr.u-tokyo.ac.jp/$\sim$mawatari/\\PANHIT/PANHIT.html)
          }

%% Appendix material should be preceded with a single \appendix command.
%% There should be a \section command for each appendix. Mark appendix
%% subsections with the same markup you use in the main body of the paper.

%% Each Appendix (indicated with \section) will be lettered A, B, C, etc.
%% The equation counter will reset when it encounters the \appendix
%% command and will number appendix equations (A1), (A2), etc. The
%% Figure and Table counter will not reset.

\appendix

\section{Forbidden region in $A_V$ versus $SFR$ parameter space}\label{sec:Av2SFR} 

%\begin{figure}[httb]
\begin{figure}[]
\begin{center}
\includegraphics[width=0.5\linewidth, angle=0]{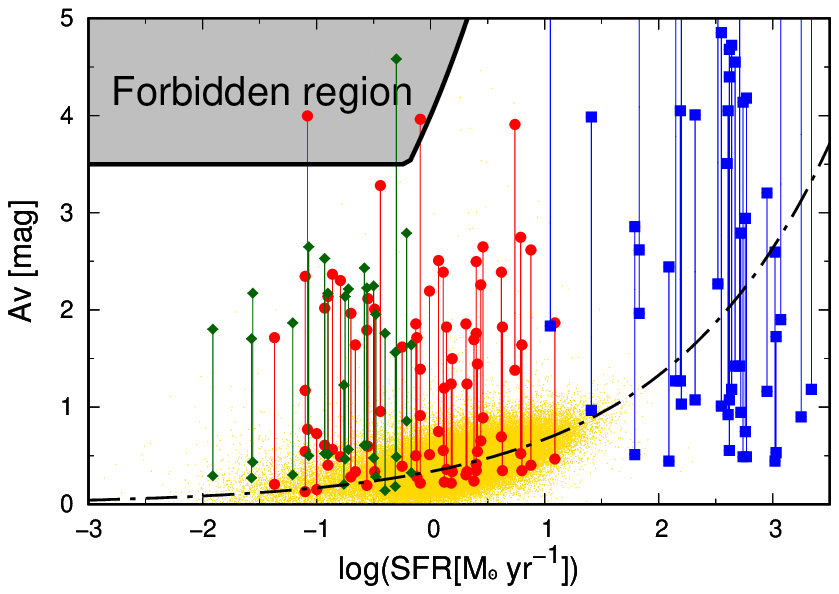}
\end{center}
\caption{Forbidden region in SFR versus $A_V$ parameter space in the SED fitting (grey shaded area). The dot-dashed curve corresponds to the relation expected from the Kennicutt-Scdmit law \citep{Kennicutt98}, gas-to-dust ratio in the Milky Way \citep{Bohlin+78,Rachford+09,Draine11}, and empirical relations of size-to-mass \citep{Shen+03} and mass-to-SFR \citep{Speagle+14} for the main sequence galaxies at $z \sim 0$. The yellow dots show the measurements for the SDSS galaxies \citep{Abazajian+09,Chen+10}. Blue squares, red circles, and green diamonds are the
$Herschel$-detected sample of SMGs at $z > 1$ \citep{Magnelli+12,Rowlands+14}, dusty galaxies with early-type morphology \citep{Rowlands+12}, and dusty passive galaxies with late-type morphology \citep{Rowlands+12}, respectively. For every $Herschel$-detected galaxy, we show two kinds of dust attenuation, $A_V^{\rm BC}$ and $A_V^{\rm ISM}$, that are connected with vertical lines. Dust attenuation for stellar emission from a whole individual galaxy should be between the $A_V^{\rm BC}$ and $A_V^{\rm ISM}$. \label{fig:Av2SFR}}
\end{figure}

% Data
There is a good correlation between $A_V$ and SFR observed so far \citep{Sullivan+01,Garn+10a,Garn+10b,Price+14}. 
% Sugahara analysis on SDSS
We compiled galaxies in the literature to examine their distribution in the SFR--$A_V$ plane. For the Sloan Digital Sky Survey (SDSS) DR7 sample \citep{Abazajian+09}, we used $\sim 100,000$ galaxies for which the H$\alpha$ and H$\beta$ emission line fluxes are available in the MPA-JHU catalog\footnote{https://wwwmpa.mpa-garching.mpg.de/SDSS/DR7/}. The SFR is estimated from the H$\alpha$ line flux \citep{Brinchmann+04}. The $A_V$ is evaluated from the H$\alpha$/H$\beta$ flux ratio \citep{Chen+10} followed by a conversion from nebular to stellar attenuation ($\times 0.44$; \citealt{Calzetti+00}). We confirmed a correlation between the $A_V$ and SFR for the local galaxies (Figure~\ref{fig:Av2SFR}). 
%\citet{Garn+10b} analyze $\sim 90,000$ local galaxies with the H$\alpha$ emission line from the Sloan Digital Sky Survey (SDSS). As a result, they find no galaxy with an SFR~$< 0.1$\,M$_\odot$\,yr$^{-1}$ and an $A_{{\rm H}\alpha} > 3$, where $A_{{\rm H}\alpha}$ is attenuation for the H$\alpha$ emission line.  

% Herschel sample
We also investigated $Herschel$-detected samples comprising 29 submillimeter galaxies (SMG) at $z > 1$ \citep{Magnelli+12,Rowlands+14}, 42 galaxies at $z < 0.5$ with early-type morphology \citep{Rowlands+12}, and 19 galaxies at $z < 0.5$ with small SFR and late-type morphology \citep{Rowlands+12}. They are rare populations of galaxies and supplementary to local typical galaxies from the SDSS sample. \citet{Rowlands+12,Rowlands+14} released physical quantities of the $Herschel$-detected dusty galaxies estimated with an SED fitting code MAGPHYS \citep{daCunha+08}, from which we extracted values of the SFR and dust optical depth for stellar emission. There are two types of dust optical depths: one of stellar birth clouds (BC) and the other of the interstellar medium (ISM) because MAGPHYS allows different amounts of attenuation in the BC and ISM \citep{daCunha+08}. We show both attenuation values in the BC and ISM ($A_V^{\rm BC}$ and $A_V^{\rm ISM}$) for $Herschel$-detected dusty galaxies in Figure~\ref{fig:Av2SFR}. The $A_V^{\rm BC}$ is always larger than $A_V^{\rm ISM}$ \citep{Charlot+00,daCunha+08}. The dust attenuation for stellar emission averaged over the whole galaxy is expected to be between $A_V^{\rm BC}$ and $A_V^{\rm ISM}$, depending on the luminosity ratio of young stars in the BC and old stars in the ISM \citep{daCunha+08}. 
%We estimated the upper limit on $A_V$ as $A_V^{\rm upper} = {\rm min} [ 2.5 \log \{1 - 1/f_{\mu} + \exp(\tau_V^{\rm ISM}) /f_\mu\}, 1.086 \tau_V^{BC} ]$, where $\tau_V^{\rm ISM}$ is the dust optical depth in the ISM (i.e., the optical depth in stellar birth clouds is not included; \citealt{daCunha+08}) and $f_\mu$ is the fraction of the ISM dust emission in the total infrared luminosity (i.e., $1 - f_\mu$ is the fraction of the luminosity from the birth clouds; \citealt{daCunha+08}). 

% Physical interpretation => Av-SFR restrinction
We can physically interpret the correlation between $A_V$ and SFR (Figure~\ref{fig:Av2SFR}) via a well-established correlation between gas column density and surface SFR density (Schmidt-Kennicutt law; \citealt{Schmidt59,Kennicutt98}). Given a dust-to-gas ratio from a Milky Way measurement ($A_V / N_{\rm H} = 5.3 \times 10^{-22}$\,cm$^2$\,mag\,atms$^{-1}$; \citealt{Bohlin+78,Rachford+09,Draine11}), the Schmidt-Kennicutt law leads to a correlation between surface SFR density and dust attenuation. We convert the surface SFR density to SFR by combining a size-mass relation \citep{Shen+03} and a mass-SFR relation (main sequence of star-forming galaxies; \citealt{Speagle+14}). In Figure~\ref{fig:Av2SFR}, we show the expected relation, $A_V = 0.9 \times {\rm SFR}^{0.3}$, by the dot-dashed curve which well reproduces the measurements for the SDSS galaxies. 

% SF threshold
%If there is a minimum gas surface density for star-formation \citep{Kennicutt89}, a galaxy with zero star-formation may still have a gas surface density below the threshold. If the gas contains dust, a certain amount of dust attenuation can be present. A possible dust attenuation for no star-formation is at most $A_V = 0.3$, assuming the dust-to-gas ratio of $A_V/N_{\rm H} = 5.3 \times 10^{-22}$\,cm$^2$\,mag\,atms$^{-1}$ and a critical gas surface density for star-formation of $5\,M_{\odot}$\,pc$^{-2}$ \citep{Kennicutt89}. 

% threshold in this work
Based on the above consideration, we conservatively define a forbidden region in the $A_V$--SFR plane for this work as $A_V > \rm{max}(4 \times SFR^{0.3}, 3.5)$ (shaded area in Figure~\ref{fig:Av2SFR}). There are three $Herschel$-detected galaxies whose $A_V^{\rm BC}$ and SFR are inside the forbidden region. We checked fractions of dust attenuated energy by the BC and ISM ($f_{\mu}$ in \citealt{Rowlands+12}), confirming that dust attenuation in the three galaxies largely occurs in the ISM. Then, the dust attenuation for whole stellar emission from the individual galaxies should be similar to the $A_V^{\rm ISM}$, which is far from the forbidden region.

\section{BBG candidates with ALMA detection}\label{sec:Fit4ALMAcontami}

%\begin{figure}[httb]
\begin{figure}[]
\begin{center}
\includegraphics[width=0.5\linewidth, angle=0]{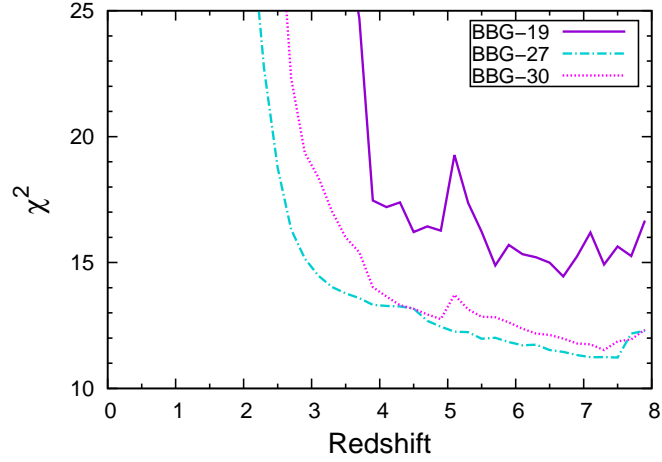}
\end{center}
\caption{SED fitting $\chi^2$ as a function of redshift for the three BBG candidates with ALMA detection. \label{fig:chi2_ALMAcontami}}
\end{figure}

%\begin{figure}[httb]
\begin{figure*}[]
\begin{center}
\includegraphics[width=1.0\linewidth, angle=0]{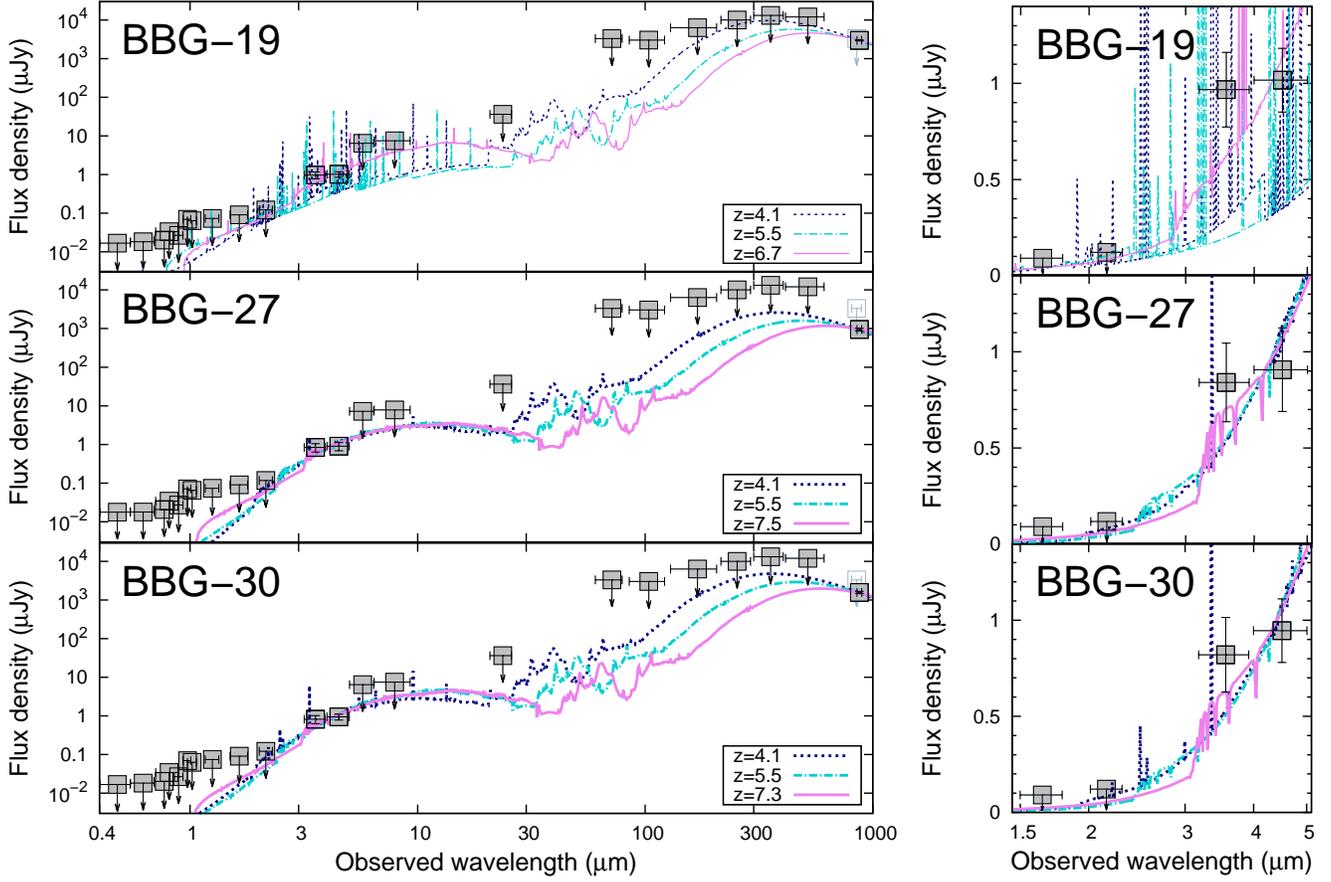}
\end{center}
\caption{Observed SEDs of the BBG candidates with ALMA detection are shown, where filled (open) squares correspond to the observed photometry used (excluded) in the template fitting. For the non-detection bands, the $2\,\sigma$ limiting fluxes are set as upper limits, indicated by arrows. The DG/DNLE model spectra at $z \sim 4$ (dotted), $5.5$ (dot-dashed), and $7$ (solid) are superposed. All the three model spectra can reasonably fit the observed SED, with the $z \sim 7$ model being the best-fit templates. The right panels are same as the left ones but in the NIR wavelength range. \label{fig:ALMAcontami_SED}}
\end{figure*}

For the three BBG candidates detected in the ALMA observations, we performed SED fitting in the same manner as in \S\ref{sec:fitmethod}. In the FIR regime, not only the ALMA Band\,7 flux density but also the upper limits in other instruments were useful to constrain the overall IR SED shape. Therefore, we used all bands between $0.4\,\mu$m and $1000\,\mu$m, except for the SCUBA-2 whose wavelength is almost the same as that of the ALMA Band~7. 

We found that DG and DNLE templates in the Galaxy group are significantly preferred to the AGN group templates for all the three objects. Their physical properties are not constrained very much, except for dust attenuation (and the IR luminosity as its deriviative). This is due to the large redshift uncertainties. The $\chi^2$ values are roughly constant at $z \gtrsim 4$, as shown in Figure~\ref{fig:chi2_ALMAcontami}. 
%While the best-fit models are at $z \sim 7$ for all of the three objects, we cannot rule out other redshifts between $4 \lesssim z \lesssim 8$. 
In contrast, 300 MC runs result in relatively narrow ranges of IR luminosity of $12 \lesssim \log(L_{\rm IR}/L_{\odot}) \lesssim 12.6$ because of the so-called ``negative K correction'' (e.g., \citealt{Blain+02}).  Figure~\ref{fig:ALMAcontami_SED} shows some DG and DNLE model spectra at different redshifts that reasonably fit the observed SEDs of the three ALMA detected galaxies. 
%While the XzLBG group templates also reproduce the observed SEDs well, they can be recognized as the high-$z$ extremes of what we obtain with the Galaxy group templates. 
We conclude that the three BBG candidates with ALMA detections are actually ultra luminous infrared galaxies with $L_{\rm IR} > 10^{12}\,L_{\odot}$ (ULIRGs; \citealt{Lonsdale+06}) at $z \gtrsim 4$. Such massive dusty populations have been recently reported by \citet{Wang+19}. Future redshift confirmation with deep spectroscopy (e.g., by $JWST$ or ALMA) is required to further constrain their physical properties.

%% The reference list follows the main body and any appendices.
%% Use LaTeX's thebibliography environment to mark up your reference list.
%% Note \begin{thebibliography} is followed by an empty set of
%% curly braces.  If you forget this, LaTeX will generate the error
%% "Perhaps a missing \item?".
%%
%% thebibliography produces citations in the text using \bibitem-\cite
%% cross-referencing. Each reference is preceded by a
%% \bibitem command that defines in curly braces the KEY that corresponds
%% to the KEY in the \cite commands (see the first section above).
%% Make sure that you provide a unique KEY for every \bibitem or else the
%% paper will not LaTeX. The square brackets should contain
%% the citation text that LaTeX will insert in
%% place of the \cite commands.

%% We have used macros to produce journal name abbreviations.
%% \aastex provides a number of these for the more frequently-cited journals.
%% See the Author Guide for a list of them.

%% Note that the style of the \bibitem labels (in []) is slightly
%% different from previous examples.  The natbib system solves a host
%% of citation expression problems, but it is necessary to clearly
%% delimit the year from the author name used in the citation.
%% See the natbib documentation for more details and options.

\bibliographystyle{aasjournal}
%\bibliography{example} % if your bibtex file is called example.bib
\bibliography{Paper_Ref} % if your bibtex file is called example.bib

%% This command is needed to show the entire author+affilation list when
%% the collaboration and author truncation commands are used.  It has to
%% go at the end of the manuscript.
%\allauthors

%% Include this line if you are using the \added, \replaced, \deleted
%% commands to see a summary list of all changes at the end of the article.
%\listofchanges

\end{document}